\documentstyle[12pt]{article}
\newcounter{c1} \newcounter{c2}
\newenvironment{eqn}{\setcounter{c1}{\value{equation}}
\setcounter{c2}{0}\addtocounter{c1}{1}
\renewcommand{\theequation}{A.\arabic{c1}\alph{c2}}
\begin{eqnarray}}{\end{eqnarray}\setcounter{equation}{\value{c1}}
\renewcommand{\theequation}{A.\arabic{equation}}}

\newcommand{\aum}{\addtocounter{c2}{1}}

\renewcommand{\theequation}{\thesection.\arabic{equation}}
\newcommand{\bb}{\begin{equation}}
\newcommand{\ee}{\end{equation}}
\newcommand{\p}{\partial}
\newcommand{\m}{\mbox{$\frac{1}{2}$}}
\newcommand{\br}{\begin{eqnarray}}
\newcommand{\er}{\end{eqnarray}}

\begin{document}
{}~\\
{}~\\
{}~\\
{}~\\
\begin{center}
{\huge {\bf Geometry of the 2+1 Black Hole}}\\
\vspace{3cm}
{\large M\'aximo Ba\~nados$^{1,2,*}$, Marc Henneaux$^{1,3,\#}$}, \\
{\large Claudio Teitelboim$^{1,2,4,*}$ and Jorge Zanelli$^{1,2,*}$} \\
\vspace{1.5cm}
{}~$^1${\em Centro de Estudios Cient\'{\i}ficos de Santiago
Casilla 16443, Santiago 9, Chile} \\
{}~$^2${\em Facultad de Ciencias, Universidad de Chile, Casilla 653,
Santiago, Chile.} \\
{}~$^3$ {\em Facult\'e des Sciences, Universit\'e
Libre de Bruxelles, Belgium.} \\
{}~$^4$ {\em Institute for Advanced Study, Olden Lane,
Princeton, New Jersey 08540, USA. }\\
\vspace{5cm}
Submitted to Phys. Rev. D. \\
November 1992
\end{center}

\setcounter{page} 0
\newpage

\begin{abstract}

The geometry of the spinning black holes of standard Einstein
theory in 2+1 dimensions, with a negative cosmological constant
and without couplings to matter, is analyzed in detail. It is
shown that the black hole arises from identifications of points
of anti-de Sitter space by a discrete subgroup of $SO(2,2)$. The
generic black hole is a smooth manifold in the metric sense. The
surface $r=0$ is not a curvature singularity but, rather, a
singularity in the causal structure.  Continuing past it would
introduce closed timelike lines. However, simple examples show
the regularity of the metric at $r=0$ to be unstable: couplings
to matter bring in a curvature singularity there. Kruskal
coordinates and Penrose diagrams are exhibited. Special
attention is given to the limiting cases of (i) the spinless
hole of zero mass, which differs from anti-de Sitter space and
plays the role of the vacuum, and (ii) the spinning hole of
maximal angular momentum .  A thorough classification of the
elements of the Lie algebra of $SO(2,2)$ is given in an
Appendix.
\smallskip
\bigskip

\noindent{PACS numbers 04.20 Jb, 97-60. Lf.}

\end{abstract}

\bigskip
\bigskip
\vfill\break

\newpage
\section{Introduction}
\setcounter{equation} 0

The black hole is one of the most fascinating structures that
has ever emerged out of the theory of gravitation. And yet, it would
seem fair to say, we are far from fully understanding
it. It is therefore fortunate that full-fledged black holes have
been found to exist\cite{1} in the transparent setting of 2+1
standard Einstein gravity\cite{2}.

The purpose of this article is to study in detail the geometry
of the 2+1 black hole without electric charge\cite{3}. These results on the
black
 hole geometry were only announced and briefly summarized in\cite{1}.

The plan of the article is the
following: Section 2 deals with the action principle and its
Hamiltonian version.  The Hamiltonian is specialized to the case
of axially symmetric time independent fields and the equations
of motion are solved. The resulting metric has two integration
constants which are next identified as the mass and angular
momentum. This identification is achieved through an analysis of
the surface integrals at spacelike infinity that must be added
to the Hamiltonian in order to make it well defined. It is then
shown that for a certain range of values of the mass and angular
momentum the solution is a black hole. This black hole is shown
to be quite similar to its 3+1 counterpart -the Kerr solution.
It has an ergosphere and an upper bound in angular momentum for
any given mass.

The discussion of section 2 focuses on the physical properties of
the black hole and ignores a question that must have been
needling the geometer hiding within every theorist. The
spacetime geometry of the black hole is one of constant negative
curvature and therefore it is, locally, that of anti-de Sitter
space. Thus, the black hole can only differ from anti-de Sitter
space in its global properties. More precisely, as we shall see, the
black hole arises from anti-de Sitter space through
identifications of points of the latter by means of a discrete
subgroup of its symmetry group\cite{4}.
Section 3 is devoted to this issue. The
identifications are explicitly given and are, in particular,
used to show that the black hole singularity at $r = 0$ is
not one in the metric, which is regular there, but rather a singularity in
the causal structure.  Continuing past $r = 0$ would bring in closed
timelike lines. When there is no angular momentum an additional
pathology appears at $r=0$, a singularity in the manifold
structure of the type present in the Taub-NUT space. This is
dealt with in Appendix B.

Once the identifications are geometrically understood, we pass,
in Section 4, to exhibit special coordinate systems which
reveal the causal structure. In particular, Kruskal coordinates are
defined and the Penrose diagrams are drawn. Special issues
pertaining to the extreme rotating black hole with non-zero mass
and to the zero mass limit of a non-rotating hole ( ``vacuum")
are analyzed. Section 5 is devoted to some concluding remarks, showing the
 instability of the regularity of the metric at $r^2 = 0$ in the presence of
 matter. It is also briefly discussed how ``chronology is protected" in
the 2+1 black hole.

The classification of the elements of the Lie algebra of the symmetry group
$SO(2,2)$ is given in Appendix A.


\section{Action Principle, Equations of Motion
\newline and their Solutions. }
\setcounter{equation} 0


\subsection{Action Principle}

The action in lagrangian form may be taken to be

\bb
I= \frac{1}{2\pi} \int \sqrt{-g} \left[ R + 2l^{-2}
\right]d^2 xdt  + B',
\label{1}
\ee
where $B'$ is a surface term and the radius $l$ is related to the
cosmological constant by $-\Lambda = l^{-2} $. [ Note that, for
convenience in what follows, the numerical factor $(16 \pi G)^{-1}$ in
front of
the action is taken to be $(2 \pi) ^{-1}$, i.e., we set the gravitational
 constant $G$, which has the dimensions of an inverse energy, equal to
 $\frac{1}{8}$].

Extremization of the action with respect to the spacetime metric
$g_{\mu \nu}(x,t)$, yields the Einstein field equations

\bb
R_{\mu \nu} - \frac{1}{2} g_{\mu \nu} (R + 2 l^{-2}) = 0
\label{2}
\ee
which, in a three dimensional spacetime, determine the full
Riemann tensor as

\bb
R_{\mu \nu \lambda \rho} = -l^{-2} (g_{\mu \lambda} g_{\nu \rho}
- g_{\nu \lambda} g_{\mu \rho})
\label{3}
\ee
describing a symmetric space of constant negative curvature.

One may pass to the hamiltonian form of (\ref{1}), which reads

\bb
I = \int\left[ \pi^{ij}\dot{g}_{ij} - N^{\perp}{\cal H}_{\perp} -N^i{\cal
H}_i \right] d^2xdt + B
\label{4}
\ee

The surface term $B$ will be discussed below. It differs from
the $B'$ appearing in the lagrangian form because the
corresponding volume integrals differ by a surface term. The
surface deformation generators ${\cal H}_{\perp} $ ,$
{\cal H}_i$ are given by

\br
{\cal H}_{\perp} & = & 2\pi g^{-1/2}(\pi^{ij}\pi_{ij}-
(\pi^i_i)^2)- (2\pi)^{-1}g^{1/2}(R+2/l^2)  \\
{\cal H}_i & = & -2\pi^j_{i/j}
\er

Extremizing the hamiltonian action with respect to the the lapse
and shift functions $ N^{\perp} $, $N^{i}$, yields the
constraint equations ${\cal H }_{\perp} = 0$ and ${\cal H} _i =
0$ which are the $\perp,\perp$ and $\perp,i$ components of
(\ref{2}). Extremization with respect to the spatial metric
$g_{ij}$ and its conjugate momentum $\pi^{ij}$, yields the
purely spatial part of the second order field equations
(\ref{2}), rewritten as a hamiltonian system of first order in
time.


\subsection{ Axially  symmetric stationary field}

One may restrict the action principle to a class of fields that
possess a rotational Killing vector $\p/ \p\phi$ and
a timelike Killing vector $\p /\p t$.  If the radial
coordinate is properly adjusted, the line element may be written
as

\br
ds^2 & = & -(N^{\perp})^2(r)dt^2 +
f^{-2}(r)dr^2  + r^2(N^\phi (r)dt + d\phi)^2 \nonumber \\
 & &  0 \leq  \phi < 2\pi ,\;\;  \;\; t_1 \leq t \leq t_2
\label{7}
\er

The form of the momenta  $\pi^{ij}$ may be obtained from
(\ref{7}) through their relation $\pi^{ij} = - (1/2 \pi) g^{-1/2}
(K^{ij} - Kg^{ij})$ with the extrinsic curvature $K_{ij}$,
which, for a time-independent metric, simply reads $2 N^{\perp} K_{ij} =
(N_{i|j} + N_{j|i})$. This gives as the only component of the
momentum,

\bb
\pi^r\,_{\phi} = \frac{l}{2 \pi} p(r)
\label{8}
\ee

If expressions (\ref{7}), (\ref{8}) are introduced in the action, one finds

\bb
I= -(t_2 -t_1)\int dr\left[ N(r){\cal H}(r) +N^{\phi}{\cal
H}_{\phi} \right] + B
\label{9}
\ee
with

\br
 {\cal H} & \equiv & 2 \pi f(r){\cal H}_{\perp} = 2l^2 \frac{p^2}{r^3} +
(f^2)' -2\frac{r}{l^2} \\
{\cal H}_{\phi} & = & -2 lp' \\
N(r) & = & f^{-1}N^{\perp} \label{N}
\er


\subsection{Solutions }

To find solutions under the assumptions of time independence and
axial symmetry, one must extremize the reduced action (\ref{9}).
Variation with respect to $N$ and $N^{\phi}$, yields that
the generators ${\cal H}$ and ${\cal H}_{\phi}$ must vanish.
These constraint equations are readily solved to give

\br
p & = & -\frac{J}{2l} \nonumber \\
f^2 & = & -M + \left(\frac{r}{l}\right)^2 + \frac{J^2}{4r^2}
\label{sol:p}
\er
where $M$ and $J$ are two constants of integration, which will be
identified below as the mass and angular momentum, respectively.

Variation of the action with respect to $f^2$ and $p$ yields the
equations

\br
  N'& = & 0\nonumber \\
(N^{\phi})' + \frac{2lp}{r^3} N    & = & 0
\label{14}
\er
which determine $N$ and $N^{\phi}$ as

\br
N & = & N(\infty)\nonumber \\
N^{\phi} & = & - \frac{J}{2r^2}N(\infty) + N^{\phi}(\infty)
\label{2.15}
\er

The constants of integration $N(\infty)$ and $N^{\phi}
(\infty)$ are part of the specification of the coordinate
system, which is not fully fixed by the form of the line element
(\ref{7}) (see below).


\subsection{ Surface integrals at infinity }


\subsubsection{ Quick analysis}

We will be interested in including in the variational principle
the class of fields that approach  our solution (\ref{sol:p}),
(\ref{2.15}) at spacelike infinity. This means that the action
should have an extremum under variations of $g_{ij}$ and
$\pi^{ij}$ that for large $r$ approach the variations of the
expressions (\ref{sol:p}), for any $\delta M$ and $\delta J$ and
for fixed $N(\infty)$, $N^{\phi}(\infty)$ . However, as
seen most evidently from the reduced form (\ref{9}) of the
action, upon varying $g_{ij}$ and $\pi ^{ij}$ one picks up a
surface term. That is, one finds

\br
\delta I & = & (t_2 - t_1) [N(\infty) \delta M - N^{\phi} (\infty) \delta J]
 		  + \delta B \nonumber \\  & & + (\mbox{Terms vanishing
when the equations of motion hold})
\label{17}
\er

Now, one must demand that when the equations of motion hold, the
variation of the action should be zero\cite{5}. Therefore,
the boundary term $B$ in the action must be adjusted so as to
cancel the first two terms on the right side of (\ref{17}).
Thus, we put

\bb
B= (t_2 -t_1) (-N(\infty) M + N^{\phi} (\infty) J)
\label{18}
\ee

Equation (\ref{18}) identifies $M$ as the mass and $J$ as the
angular momentum. This is because they appear as conjugates to
the asymptotic displacements $N(\infty)$ and $N^{\phi}(\infty)$.
(The minus sign in front of $N(\infty)$ appears because,
conventionally, one introduces a minus sign in the generator
when the displacement is along a timelike direction.) That
$N^{\phi}$ is the angular displacement is evident. However, the
fact that the rescaled lapse $N$ given by (\ref{N}) appears in
(\ref{18}) rather than the original $N^{\perp}$, deserves
explanation. The reason is the following.  The normal component
of the deformation that joins the surface of time $t$ and that
of time $t + \delta t$ is $\delta {\bf \xi} = {\bf n} N^{\perp}
\delta t$, where ${\bf n}$ is the unit
normal. But the unit normal does not approach a Killing vector
at infinity. If one multiplies it by $f$, one obtains, at
infinity, a Killing vector ${\bf K} = {\bf n}f$ whose norm ${\bf
K} \cdot {\bf K} = - f^2$ is independent of $N(\infty)$. The
displacement $N(\infty) \delta t$ (``Killing time") is the
component of the deformation $\delta{\bf \xi}$ along ${\bf K}$.


\subsubsection{ Detailed analysis}

The preceding argument gives a quick way of obtaining the
surface integrals that must be added to the action. It also puts
in evidence the physical meaning of $M$ and $J$. However, a
more careful analysis is needed. One knows that in a gauge theory such as
 General
Relativity the conserved quantities are related to the
asymptotic symmetry group. This fact already
emerged in the previous discussion where ``displacements at
infinity" played the key role. For 2+1 spacetime dimensions with
a negative cosmological constant, this asymptotic group is
infinite dimensional and contains $SO(2,2)$ as a subgroup. The
asymptotic Killing vectors $\p /\p{\phi}$ and ${\bf K}  =
N(\infty)^{-1}\p_t$ that appeared above are two of the
generators in the Lie algebra of $SO(2,2)$. Thus, what we have
called ``Killing time displacements" are not ``translations"
but -rather- $SO(2,2)$ boosts.

The general analysis of the asymptotic symmetry group
of 2+1 gravity has been given in \cite{6}. We briefly recall
here its key aspects and apply them to the present treatment.

One considers all metrics that for large $r$ become

\bb
ds^2 \longrightarrow -\left(\frac{r}{l}\right)^2 dt^2 +
\left(\frac{r}{l}\right)^2 dr^2 + r^2 d\phi ^2,
\label{19}
\ee
[There is no loss of generality -in this context- in taking
$N(\infty) = 1$ and $N^{\phi}(\infty) =0$. One must only
remember that for any given spacetime the surface integrals are
to be calculated in a coordinate system obeying these conditions.]

The precise way in which $ds^2$ approaches (\ref{19}) for large
$r$ is obtained by acting on the solution (\ref{sol:p}),
(\ref{2.15}), with all possible anti-de Sitter group
transformations. The rationale for this procedure is that one
wants to have \underline{at least} $SO(2,2)$ as an asymptotic
symmetry group. This is because the metric (\ref{19}) coincides
with the asymptotic form of the anti-de Sitter metric, which has
$SO(2,2)$ as its (exact) symmetry group.  The remarkable feature
is that the resulting class of allowed asymptotic metrics admits
a much larger symmetry group.

The asymptotic symmetry group turns out to be the conformal
group. The conformal group may be defined as the group of all
transformations that leave invariant the cylinder at infinity,
up to a Weyl rescaling. The conformal Killing vectors obey

\bb
\xi_{\alpha ;\beta} + \xi_{\beta ;\alpha} -\frac{1}{2} g_{\alpha
\beta} \xi^{\lambda}_{\; ;\lambda} = 0
\label{20}\ee

The Lie algebra of the conformal group consists of two copies of
the Virasoro algebra. Therefore the conserved charges of 2+1
gravity are two sets $L_n$, $K_n$ of Virasoro generators $(n =
0, \pm 1, \pm 2, ....)$. Of these, the six $SO(2,2)$ generators
are $ L_0, L_1, L_{-1}, K_0, K_1, K_{-1}$, which form a
subalgebra .

The $L_n$ and $K_n$ obey the Virasoro algebra with a central
charge proportional to the radius of curvature. One has, in
terms of non-quantum Poisson brackets,

\br
{[L_{n}, L_{m}]} &=& -i\{(n-m)L_{n+m} + l\cdot n(n^2-1)\delta_{n, -m} \}
\nonumber \\
{[K_{n}, K_{m}]} &=& -i\{(n-m)K_{n+m} + l\cdot n(n^2
-1)\delta_{n,-m} \}  \label{2.20.1} \\
{[L_{n}, K_{m}]} &=& 0  \nonumber
\er

In the normalization for the central charge that has become
standard in string theory, one has

\bb
c= 12l/\hbar
\label{2.20.2}
\ee

The metric given by (\ref{sol:p}), (\ref{2.15}) has only two
charges which are non-zero ($M =K_0 +L_0,\;\; J= K_0 -L_0$).
However, by acting with the asymptotic group one can endow it
with other charges, much as by boosting a Schwarzschild solution
one may endow it with linear momentum.


\subsection{The Black Hole}

The lapse function $N^{\perp}$ vanishes for two values of $r$
given by

\bb
r_{\pm}=l \left[ \frac{M}{2} \left( 1 \pm \sqrt{1-\left(
\frac{J}{Ml}\right)^2} \right) \right]^{1/2}.
\label{21}
\ee
whereas $g_{00}$ vanishes at

\bb
r_{erg} = l M^{1/2}
\label{22}
\ee

These three special values of $r$ obey

\bb
r_{-} \leq r_{+} \leq r_{erg}
\label{23}
\ee

Just as it happens in 3+1 dimensions for the Kerr metric, $r_{+}$
is the black hole horizon, $r_{erg}$ is the surface of infinite
redshift and the region between $r_{+}$ and $r_{erg}$ is the
ergosphere. In order for the solution to describe a black hole,
one must have

\bb
    M>0,  \;\;\;\;  |J| \leq Ml.
\label{24}
\ee
In the extreme case $|J|=Ml$, both roots of $N^2=0$ coincide.
Note that the radius of curvature $l=(-\Lambda)^{-1/2}$ provides the
length scale necessary in order to have a horizon in a theory in which the
mass is dimensionless. If one lets $l$ grow very large the black hole
exterior is pushed away to infinity and one is left just with the inside.

The vacuum state, namely what is to be regarded as empty space, is
obtained by making the black hole disappear. That is, by letting
the horizon size go to zero. This amounts to letting
$M\rightarrow 0$, which requires $J\rightarrow 0$ on account of
(\ref{24}). One thus obtains the line element

\bb
ds^2_{vac} = -(r/l)^2 dt^2 + (r/l)^{-2} dr^2 + r^2 d\phi^2.
\label{25}
\ee

As $M$ grows negative one encounters the solutions studied
previously in \cite{7}. The conical singularity that
they possess is naked, just as the curvature singularity of a
negative mass black hole in $3+1$ dimensions.  Thus, they must,
in the present context, be excluded from the physical spectrum.
There is however an important exceptional case. When one reaches
$M=-1$ and $J=0$ the singularity disappears.  There is no
horizon, but there is no singularity to hide either.  The
configuration

\bb
 ds^2=-(1+(r/l)^2) dt^2 + (1+(r/l)^2)^{-1} dr^2 + r^2 d\phi^2
\label{26}
\ee
(anti-de Sitter space) is again permissible.

Therefore, one sees that anti-de Sitter space emerges as a ``bound state",
separated from the continuous black hole spectrum by a mass gap of one
unit. This state cannot be deformed continuously into the vacuum
(\ref{25}), because the deformation would require going through a sequence
of naked singularities which are not included in the configuration space.

Note that the zero point of energy has been set so that the mass vanishes
when the horizon size goes to zero. This is quite natural. It is what is
done in 3+1 dimensions. In the past, the zero of energy has been adjusted
so that anti-de Sitter space has zero mass instead. Quite apart from
this difference, the key point is that the black hole spectrum lies above
the limiting case $M=0$.

We now pass, in the next section, to a detailed study of the
geometry of the black hole.


\section{Black Hole as Anti-de Sitter Space Factored by a
Subgroup of its Symmetry Group}
\setcounter{equation} 0

We will show in this section that the black hole arises from
anti-de Sitter space through identifications by means of a
discrete subgroup of its isometry group $SO(2,2)$. This implies
that the black hole is a solution of the source-free Einstein
equations everywhere, including $r=0$. As we shall also see, the
type of ``singularity" that is found at $r=0$ is -generically-
one in the causal structure and not in the curvature, which is
everywhere finite (and constant).  It should be emphasized that
this statement means that $r = 0$ is not
a conical singularity.

To proceed with the analysis we first review the properties of
anti-de Sitter space.

\subsection{Anti-de Sitter Space in 2+1 Dimensions}


\subsubsection{Metric}

Anti-de Sitter space can be defined in terms of its embedding in a
four dimensional flat space of signature $(- -++)$

\bb
ds^2=-du^2-dv^2+dx^2+dy^2
\label{3.1}
\ee
through the equation

\bb
-v^2-u^2+x^2+y^2=-l^2.
\label{3.2}
\ee

A system of coordinates covering the whole of the manifold may be
introduced by setting

\bb
u=l\cosh\mu \sin\lambda, \;\;\; v=l\cosh\mu \cos\lambda
\label{3.3}
\ee
with $l\sinh\mu=\sqrt{x^2+y^2}$ and $0\leq \mu <\infty$,
$0\leq\lambda <2\pi$.  Inserting (\ref{3.3}) into (\ref{3.1})
gives

\bb
ds^2= l^2\left(- \cosh^2\mu \ d\lambda^2 + \frac{dx^2+dy^2 }{l^2 +x^2
+y^2}\right)
\label{3.4}
\ee
an expression that can be further simplified by passing to polar
coordinates in the $x-y$ plane

\bb
x=l\sinh\mu \cos \theta, \;\;\; y=l\sinh \mu \cos \theta,
\label{3.5}
\ee
which yields

\bb
ds^2= l^2\left[ -\cosh^2 \mu d\lambda^2 + d \mu^2 + \sinh^2 \mu
d\theta^2 \right]
\label{3.6}
\ee
for the metric of anti-de Sitter space.

Because $\lambda$ is an angle, there are closed timelike curves
in anti-de Sitter space (for instance $\mu=\mu_{0},
\theta =\theta_{0}$). For this reason, one ``unwraps" the $\lambda$
coordinate, that is, one does not identify $\lambda$ with
$\lambda+2\pi$. The space thus obtained is the universal
covering of anti-de Sitter space. It is this space which, by a
common abuse of language, will be called anti-de Sitter space in
the sequel. If the unwrapped $\lambda$ is denoted by $t/l$ and
if one sets $r=l\sinh\mu$, one obtains

\bb
ds^2= ((r/l)^2+1)dt^2 + ((r/l)^2+1)^{-1}dr^2 +r^2d\theta^2
\label{3.6.5}
\ee
which is the metric (\ref{7}) with $M=-1$, $J=0$ (and
$\phi$ replaced by $\theta$).


\subsubsection{Isometries}

By construction, the anti-de Sitter metric is invariant under
$SO(2,2)$. The Killing vectors are

\bb
J_{a b}=x_b \frac{\p}{\p x^a} - x_a
\frac{\p}{\p x^b}
\label{3.7}
\ee
where $x^a=(v,u,x,y)$ or, in detail

\bb
\begin{array}{rclcrcl}
J_{01} &=& v\p_u - u \p_v \;\;&\; J_{02} &=& x\p_v
+ v \p_x \\
J_{03} &=& y\p_v + v \p_y \;\;&\; J_{12} &=& x\p_u
+ u \p_x \\
J_{13} &=& y\p_u + u \p_y \;\;&\; J_{23} &=& y\p_x
- x \p_y
\end{array}
\label{3.8}
\ee

The vector $J_{01}$ generates ``time displacements"
($J_{01}=\p_{\lambda}$) whereas $J_{23}$ generates
rotations in the $x-y$ plane ($J_{23}=\p_{\theta}$). The
most general Killing vector is given by

\bb
\m \omega^{ab} J_{ab},\;\; \;\;\; \omega^{ab}=-\omega^{ba}
\label{3.9}
\ee
and is thus determined by an antisymmetric tensor in ${\bf R}^4$.


\subsubsection{Poincar\'e Coordinates}

The coordinates defined by

\bb
z=\frac{l}{u+x}, \;\;\; \beta=\frac{y}{u+x}, \;\;\;
\gamma=\frac{-v}{u+x}.
\label{3.10}
\ee
are called Poincar\'e coordinates. They only cover part of the
space, namely just one of the infinitely many regions where $u+x$
has a definite sign (see Fig. 1).  These coordinates are
therefore not well adapted to the study of global properties. In
terms of $(z,\beta,\gamma)$ the anti-de Sitter line element
reads

\bb
ds^2 = l^2 \left[ \frac{ dz^2 + d\beta^2 - d\gamma^2 }{z^2}
\right].
\label{3.11}
\ee
For $u+x>0$ one has $z>0$ and for $u+x<0$ one has $z<0$. One can
also find analogous Poincar\'e coordinates for each of the regions
where $u-x$ has a definite sign.


\subsection{Identifications}


\subsubsection{Identification subgroup associated with a Killing vector}

Any Killing vector $\xi$ defines a one parameter subgroup of
isometries of anti-de Sitter space

\bb
P \rightarrow e^{t\xi} P.
\label{3.12}
\ee
The mappings of (\ref{3.12}) for which $t$ is an integer
multiple of a basic ``step", taken conventionally as $2\pi$,

\bb
P\rightarrow e^{t\xi}P, \;\;\;\;\;  t=0,\pm 2\pi,\pm 4\pi,....
\label{3.13}
\ee
define what we will call the identification subgroup

Since the transformations (\ref{3.13}) are isometries, the
quotient space obtained by identifying points that belong to a
given orbit of the identification subgroup, inherits from
anti-de Sitter space a well defined metric which has constant
negative curvature. The quotient space thus remains a solution
of the Einstein equations.

The identification process makes the curves joining two points
of anti-de Sitter space that are on the same orbit to be closed
in the quotient space. In order for the quotient space to have an
admissible causal structure, these new closed curves should not
be timelike or null. A necessary condition for the absence of closed
timelike lines is that the Killing vector $\xi$ be spacelike,

\bb
\xi \cdot \xi >0
\label{3.14}
\ee
This condition is not sufficient in general. However, as it
will be shown in Sec. 3.2.5, it turns out to be so for the
particular Killing vectors employed in the identifications
leading to the black hole.


\subsubsection{Singularities in the causal structure}

There are some Killing vectors that do fulfill (\ref{3.14})
everywhere in anti-de Sitter space, for example $\frac{\p}{\p
\theta}$, where $\theta$ is the angular coordinate appearing in
(\ref{3.6}).

However, the Killing vectors appearing in the identifications
that give rise to the black hole are timelike or null in some
regions.  These regions must be cut out from anti-de Sitter
space to make the identifications permissible. The resulting
space -which we denote (adS)'- is invariant under (\ref{3.12})
because the norm of a Killing vector is constant along its
orbits. Hence, the quotient can still be taken.

The space (adS)' is geodesically incomplete since one can find
geodesics that go from $\xi \cdot \xi >0$ to  $\xi \cdot
\xi <0$.  From the point of view of (adS)' -i.e., prior to the
identifications- it is quite unnatural to remove the regions
where  $\xi \cdot \xi$ is not positive. However, once the
identifications are made, the frontier of the region $\xi \cdot
\xi >0$, i.e., the surface $\xi \cdot \xi =0$, appears as a
singularity in the causal structure of spacetime, since
continuing beyond it would produce closed timelike curves.

For this reason, the region $\xi \cdot \xi =0$ may be
regarded as a true singularity in the quotient space. If this
point of view is taken, -as it is done here- the only incomplete
geodesics are those that hit the singularity, just as in the 3+1
black hole. It should be stressed that the surface $\xi \cdot
\xi =0$ is a singularity only in the causal structure. It is not
a conical curvature singularity of the type discussed in
\cite{7}.  Indeed, the quotient space is  smooth \cite{8}.  Its
curvature tensor is everywhere regular and given by

\bb
R_{\mu \nu \lambda \rho} = -l^{-2} (g_{\mu \lambda} g_{\nu \rho}
- g_{\nu \lambda} g_{\mu \rho}).
\ee

The fundamental group of the quotient space is non trivial and
isomorphic to the identification subgroup.  The orbits of the
Killing vectors define closed curves that cannot be continuously
shrunk to a point. The  ``origin"  $\xi \cdot \xi=0$ is neither a
point nor a circle. It is a surface. The topology of $\xi \cdot
\xi =0$, and also that of the whole quotient space, can be
inferred by inspection of the Penrose diagram in Fig. 4c.  One
finds that the black hole is topologically {\bf R}$^2 \times
S^1$ and that the surface $\xi \cdot \xi=0$ has infinitely many
connected pieces, each of which is a cylinder whose circular
sections are null.  \\
\\

\subsubsection{Explicit form of the identifications}
We claim that the black hole solutions are obtained by making
identifications of the type described above by the discrete
group generated by the Killing vector
\\
\bb
\xi = \frac{r_{+}}{l}J_{12} - \frac{r_{-}}{l}J_{03} - J_{13} + J_{23}
\label{3.15}
\ee
where the $J_{ab}$ are given by (\ref{3.7}). The antisymmetric
tensor $\omega^{ab}$ defined by (\ref{3.15}) through $\xi$ =
$\frac{1}{2} \omega^{ab} J_{ab}$,  is easily verified to possess
real eigenvalues, namely, $ \pm r_{+}/l$ and $\pm r_{-}/l$. The
corresponding Casimir invariants $I_1 $=$ \omega_{a b} \omega^{a
b}$ and $I_2$ = $\frac{1}{2}\epsilon _{a b c d} \omega^{a b}
\omega^{cd}$ are

\bb
I_1 = -\frac{2}{l^2}(r_{+}^2 + r_{-}^2)= -2M,\:\;\;\;
I_2 = -\frac{4}{l^2} r_{+} r_{-} = -2\frac{|J|}{l}
\label{3.16}
\ee

According to the classification given in Appendix A the Killing
vector (\ref{3.15}) is of type {\bf I}$_b$ when $r_+\neq r_-$,
of type {\bf II}$_a$ when $r_+ =r_- \not= 0$ and of type {\bf III}$^{+}$ when
 $r_{+} = r_{-}=0 $.

To prove that the identifications by $e^{2\pi k\xi}$ yield the
black hole metric, we start by considering the non - extreme
case $r_{+}^2 - r_{-}^2 > 0$. In that case, by performing an
$SO(2,2)$ transformation, one can eliminate the last term in
(\ref{3.15}) and replace $\xi$ by the simpler expression

\bb
\xi' = \frac{r_{+}}{l}J_{12} - \frac{r_{-}}{l}J_{03}
\label{3.18}
\ee
This follows from the analysis of Appendix A, where it is shown
that any $SO(2,2)$ element with unequal real eigenvalues can be
brought into the form (\ref{3.18}) by an $SO(2,2)$
transformation. Alternatively, one may rewrite (\ref{3.15}) in
Poincar\'e coordinates as

\bb
-\xi =\frac{r_{+}}{l} \left(z\frac{\p}{\p z} + \beta
\frac{\p}{\p \beta} + \gamma \frac{\p}{\p
\gamma}\right) - \frac{r_{-}}{l}\left(\beta \frac{\p}{\p \gamma}
+ \gamma \frac{\p}{\p \beta}\right) +
\frac{\p}{\p \beta}
\label{3.17}
\ee
and observe that the shifts

\br
\beta &\rightarrow& \beta - \frac{r_{+}}{r_{+}^2 - r_{-}^2}
\label{3.19.a} \\
\gamma &\rightarrow& \gamma - \frac{r_{-}}{r_{+}^2 - r_{-}^2}
\label{3.19.b}
\er
-which are $SO(2,2)$ isometries- eliminate $\frac{\p}{\p
\beta}$ in (\ref{3.17}).

The norm of $\xi '$ is given by

\bb
\xi '\cdot \xi ' = \frac{r_{+}^2}{l^2}(u^2 - x^2) +
\frac{r_{-}^2}{l^2}( v^2 - y^2)
\label{3.20}
\ee
or, using (\ref{3.2}),

\bb
\xi '\cdot \xi '= \frac{r_{+}^2 - r_{-}^2}{l^2} (u^2 - x^2) +r_{-}^2
\ee
Accordingly, the allowed region where $\xi ' \cdot \xi '>0$ is

\bb
\frac{-r_{-}^2l^2}{r_{+}^2 - r_{-}^2} < u^2 - x^2 < \infty.
\ee

The region $\xi '\cdot \xi ' >0$ can be divided in an infinite
number of regions of three different types bounded by the null
surfaces $u^2 - x^2 =0$ or $v^2 - y^2 = l^2 - (u^2 - x^2) =0$. These regions
 are:

{\bf Regions of type I}: Smallest connected regions with $u^2 - x^2 >l^2$ and
 $y$ and $u$ of definite sign. These regions have no intersection with $y
= 0$ since this would violate $u^2 - x^2 = l^2 + y^2 - v^2 >
l^2$. These regions are called ``the outer regions". The norm of the Killing
 vector fulfills  $r_{+}^2 < \xi '\cdot \xi '<+\infty$.

{\bf Regions of type II}: Smallest connected regions with $0<u^2 - x^2 < l^2$
and $u$ and $v$ of definite sign. These regions are called ``the intermediate
 regions". The norm of the Killing vector fulfills  $r_{-}^2 < \xi '\cdot \xi
 '<r_{+}^2$.

{\bf Regions of type III}: Smallest connected regions with
 $-\frac{r_{-}^2l^2}{r_{+}^2 - r_{-}^2} <
u^2 - x^2 <0$ and $x$ and $v$ of definite sign. These regions are
called ``the inner regions" and only exist for $r_{-} \neq 0$. They
do not intersect the $x = 0$ plane. The norm of the Killing vector fulfills
$0< \xi '\cdot \xi '< r_{-}^2$

The frontiers between the various regions are lightlike surfaces
(the horizons!). Each region of type {\bf I} has one region of
type {\bf II} in its future and one in its past. For $r_{-} \neq 0$, two
 situations are found for
each region of type {\bf II}: (i) it has one region of type {\bf
II} and two regions of type {\bf I} in its future as well as one
region of type {\bf II} and two regions of type {\bf III} in its past,
or conversely (ii) the same description with {\bf I} and {\bf III}
interchanged.
 Finally,
each region of type {\bf III} has one region of type {\bf II} in
its future and another one in its past.  This is shown in
Figures (2.a,b,c).
Let us now choose three contiguous regions of types {\bf I}, {\bf
II} and {\bf III}
(one of each type). In these regions we introduce a
$(t,r,\phi)$- parametrization as follows (we assume for
definiteness $u$, $y >0$ in {\bf I}, $u$, $-v >0$ in {\bf II} and $x$, $-v
>0$ in {\bf III}).

{\bf Region  I}. $r_{+} < r$:
\br
u & = & \sqrt{A(r)} \cosh \tilde{\phi}(t,\phi)  \nonumber \\
x & = & \sqrt{A(r)} \sinh \tilde{\phi}(t,\phi)  \nonumber \\
y & = & \sqrt{B(r)} \cosh \tilde{t}(t,\phi)  \nonumber \\
v & = & \sqrt{B(r)} \sinh \tilde{t}(t,\phi)
\label{3.21.a}
\er

{\bf Region II}. $r_{-} < r < r_{+}$:
\br
u & = & \sqrt{A(r)} \cosh \tilde{\phi}(t,\phi)  \nonumber \\
x & = & \sqrt{A(r)} \sinh \tilde{\phi}(t,\phi)  \nonumber \\
y & = & -\sqrt{-B(r)}\sinh \tilde{t}(t,\phi)  \nonumber \\
v & = & -\sqrt{-B(r)} \cosh \tilde{t}(t,\phi)
\label{3.21.b}
\er

{\bf Region III}. $ 0 < r <r_{-}$:

\br
u & = & \sqrt{-A(r)} \sinh \tilde{\phi}(t,\phi)  \nonumber \\
x & = & \sqrt{-A(r)} \cosh \tilde{\phi}(t,\phi)  \nonumber \\
y & = & -\sqrt{-B(r)} \sinh \tilde{t}(t,\phi)  \nonumber \\
v & = & -\sqrt{-B(r)} \cosh \tilde{t}(t,\phi)
\label{3.21.c}
\er

In (\ref{3.21.a}), (\ref{3.21.b}) and (\ref{3.21.c}) we have set

\bb
\begin{array}{rclcl}
A(r) & = & l^2\left(\frac{r^2 - r_{-}^2}{r_{+}^2 - r_{-}^2}\right) & , &
B(r) = l^2\left(\frac{r^2- r_{+}^2}{r_{+}^2 - r_{-}^2}\right)
\label{3.18.d} \\
\tilde{t} & = & (1/l) \left(r_{+}t/l - r_{-}\phi\right) & , & \tilde{\phi}  =
 (1/l)  \left(-r_{-} t/l + r_{+}\phi \right)
\label{3.18.e}\\
\label{3.21.f}
\end{array}
\ee

In the coordinates $t,\; r,\; \phi$, the metric becomes

\bb
ds^2 = -(N^{\perp})^2dt^2 + (N^{\perp})^{-2}dr^2 + r^2(N^{\phi}
dt + d\phi)^2
\label{3.22}
\ee
with $- \infty < t < \infty$, $-\infty < \phi < \infty$ i.e., it is the black
hole metric but with $\phi$ a non-periodic coordinate. The
Killing vector $\xi '$ reads

\bb
\xi ' = \frac{\p}{\p \phi}
\label{3.23}
\ee
By making the identification

\bb
\phi \rightarrow \phi + 2k\pi,
\label{3.23.1}
\ee
one gets the black hole spacetime as claimed above.

It is clear from the construction that the coordinate system
$t,\; r,\; \phi$ does not cover the domain $\xi ' \cdot \xi '
>0$ entirely, since it only covers one region of each type. If
$r_{-} = 0$  (in which case region {\bf III} does not exist), this is only
half
 of one connected component of the
domain $\xi ' \cdot \xi ' >0$. If $r_{-} \neq 0$, each of the
regions {\bf I}, {\bf II} and {\bf III} should be repeated an
infinite number of times to completely cover the domain $\xi '
\cdot \xi ' >0$ which is now connected.  This infinite pattern
follows from the fact that one is dealing with the universal
covering space of anti-de Sitter space and this will reappear in
the Penrose diagrams given below.

It is worthwhile emphasizing that it is the identification
(\ref{3.23.1}) that makes the black hole. If one does not say
that $\phi$ is an angle, one simply has a portion of anti-de
Sitter space and the horizon is just that of an accelerated observer\cite{9}.


\subsubsection{Extreme case}

The above derivation cannot be repeated in the extreme case
$r_{+} = r_{-}$. This is because the Killing vector (\ref{3.15})
is now of a different type than (\ref{3.18}). According to the
classification given in the Appendix, when $r_{+} = r_{-}$,
(\ref{3.15}) is of type {\bf II}$_a$, while (\ref{3.18}) is of type
{\bf I}$_b$ with doubly degenerate roots. Hence, there is no
$SO(2,2)$ transformation mapping one to the other.

One can nevertheless argue that the identifications for anti-de
Sitter space generated by (\ref{3.15}) yield the extreme black
hole without exhibiting the precise coordinate transformation
that brings $\xi$ into the form $\p /\p \phi$. The argument runs
as follows.  The metric (\ref{3.22}) is regular even if one sets
$r_{+}^2 = r_{-}^2$. When $\phi$ is not identified, it describes
a portion of anti-de Sitter space for any value of $r_{+}^2 -
r_{-}^2 >0$, hence it does so also in the limit $r_+-r_-
\rightarrow 0$.  Similarly, $\p /\p \phi$ is a Killing vector
for any value of $r_{-}$ and $r_{+}$.  By continuity, its Casimir
invariants remain equal to $I_1= -2(r_{+}^2 +r_{-}^2)/l^2$ and
$I_2= - 4r_{+}r_{-}/l^2$ in the limit $r_{+} - r_{-} \rightarrow
0$. Hence, in the extreme case the vector $\p /\p \phi$ remains
type I$_b$ (with coincident roots) or becomes type {\bf
II}$_a$, since these are the only two types compatible with the
given $I_1,\; I_2$. It is the latter alternative that is
realized. Indeed, type {\bf I}$_b$ may be excluded by noticing
that the corresponding Killing vector has constant norm equal to
$r_{+}^2$, whereas $\p /\p \phi$ has a space-dependent norm
equal to $r^2$. Thus $\p /\p \phi$ must be of type II$_a$ and,
thus, equal to (\ref{3.15}) [up to a possible $SO(2,2)$
transformation that leaves the metric invariant].

The preceding argument already establishes that the black hole
is obtained from anti-de Sitter space by an identification.
However, for completeness we exhibit a change of coordinates
in terms of which the identification just makes a coordinate
periodic.  The required coordinate transformation can be
explicitly given in Poincar\'e coordinates. We start with the
case $M= r_{+}=r_{-}=0$ (``the vacuum"), which is the more
illuminating one.

For $M=0$, the region $\xi \cdot \xi >0$ splits  into disjoint
regions which are just the Poincar\'e patches $u+x>0$ or
$u+x<0$. Hence, to describe a connected domain where $\xi \cdot
\xi >0$, one can just consider a single Poincar\'e patch. In
Poincar\'e coordinates the Killing vector $\xi$ is  $ -\frac{\p}{\p
\beta}$ and hence, the identifications

\bb
\beta \rightarrow \beta + 2k\pi
\label{3.24}
\ee
in (\ref{3.11}) lead to the black hole metric with $M=0$ upon
setting $z=1/r,\;\; \beta=\phi$ and $\gamma =t$.

[Note that as depicted in Fig 1, the horizon-singularity $r=0$ are the null
 surfaces $u+x=0$
delimiting the Poincar\'e region. Because the Killing vector
$\xi$ is again spacelike on the other side of $u+x=0$, one can
continue the solution with zero mass through $r=0$ to negative
values of $r$ without encountering closed timelike curves. By
doing so one includes, however, the closed lightlike curves that
lie on the null surface $u+x=0$, as well as some singularities
in the manifold structure of the type discussed in Appendix B.]

The coordinate transformation bringing the anti-de Sitter
metric to the extreme case with $M\neq 0$ (and non-periodic
in $\phi$) is more complicated. One needs in that case more
than one Poincar\'e patch to cover the black-hole
spacetime. Actually an infinite number of sets of patches
is necessary, with each set containing one patch of each of
the four types $u+x>0$, $u+x<0$, $u-x>0$,
$u-x<0$. We merely give here that transformation in one of
the patches $u+x>0$, for $r>r_{+}$.

\br
\beta &=& \frac{1}{2}\left(\frac{T}{l} + \phi +e^{2r_{+}\phi} -
\frac{1}{2r_{+}}\right)
\label{3.25.a} \\
\gamma &=& \frac{1}{2}\left(\frac{T}{l} + \phi -e^{2r_{+}\phi} +
\frac{1}{2r_{+}}\right)
\label{3.25.b} \\
z &=& \left[\frac{1}{2r_{+}}(r^2 - r_{+}^2) \right]^{-1/2} e^{r_{+}\phi}
\label{3.25.c}
\er
where $T$ is given by

\bb
T = 2t - \frac{l^2 r_+}{r^2-r^2_+}
\label{3.25.d}
\ee
and fulfills $dT = 2dt+\frac{2r_{+}l^2r dr}{(r^2 -
r_{+}^2)^2}$. By substituting (\ref{3.25.a})-(\ref{3.25.c}) in the Poincar\'e
metric, one gets the extreme black hole metric (with $N^{\phi}$
adjusted so that $N^{\phi}(r_{+})=0$).


\subsubsection{Absence of Closed Timelike Curves}

We now complete the argument that there are no closed causal
curves in the black hole solution. That is, we show that there
is no non-spacelike, future-directed, curve lying in the region
$\xi \cdot \xi >0$ of anti-de Sitter space and joining a point
and its image generated by $exp 2\pi \xi$.

Since the surfaces $r=r_{+}$ and $r=r_{-}$ are null, a causal
curve which leaves any one of the regions of types {\bf I},
{\bf II} or {\bf III} through $r=r_{+}$ or $r=r_{-}$ can never
re-enter it. Furthermore, since the images of a point are all in
the same region as that point, it is sufficient to consider each
of these regions separately.

In each of the regions of types {\bf I}, {\bf II} or {\bf III},
the anti-de Sitter metric takes the form

\bb
ds^2 = -(N^{\perp}(r))^2 dt^2 + (N^{\perp}(r))^{-2} dr^2
+r^2(N^{\phi}dt +d\phi)^2
\ee
where $\phi$ goes from $-\infty$ to $+\infty$. Consider a
causal curve $t(\lambda)\;\; r(\lambda)$ and $\phi(\lambda)$,
where the parametrization is such that the tangent vector
$(dt/d\lambda,\; dr/d\lambda,\; d\phi/d\lambda)$ does not vanish
for any value of $\lambda$. The causal property of the curve
reads

\bb
(N^{\perp})^2\left( \frac{dt}{d\lambda}\right)^2 -(N^{\perp})^{-2}\left(
\frac{dr}{d\lambda} \right)^2
-r^2\left( N^{\phi}\frac{dt}{d\lambda} +\frac{d\phi}{d\lambda} \right)^2
\leq 0.
\label{3.33}
\ee

In order to join the point $(t_0,\; r_0,\; \phi_0)$ and $(t_0,\;
r_0,\; \phi_0 +2k\pi)$, the causal curve would have to be such
that $dt/d\lambda =0$ for some value of $\lambda$, since
$t$ comes back to its initial value. But then, if $(N^{\perp})^2 > 0$ it
follows from (\ref{3.33}) that $dr/d\lambda = d\phi/d\lambda
=0$, leading to a contradiction.  Similarly, if $(N^{\perp})^2 <0$ (region
{\bf II}), the fact that $dr/d\lambda =0$ for some value
of $\lambda$ implies $dt/d\lambda= d\phi/d\lambda =0$, and the
required contradiction.  $\Box$

It should be observed that if one were to admit the region $\xi
\cdot \xi \leq 0$ in the solution, one could leave and re-enter
the regions of type {\bf III} through the surface $\xi \cdot \xi
=0$, which is timelike for $J \neq 0$. (This is not possible
when $J=0$ because the surface $\xi \cdot \xi =0$ is then null.)
One would find that there are also closed timelike curves
passing through points in region {\bf III}. The boundary between
the region where there are no closed causal curves and the
region in which there are is then the null surface $r=r_{-}$. From the point
of view of an outside observer staying at $r>r_{+}$, the inclusion or
 non-inclusion of the region  $\xi
\cdot \xi \leq 0$  is irrelevant and cannot be probed since the surface $r =
 r_{+}$ remains in all cases an event horizon.


\subsubsection{Black Hole has only two Killing vectors}

The black hole metric was obtained in Sec. 3.2.3 under the assumption of
existence
 of two commuting Killing vectors $\partial/\partial t$ and $\partial/\partial
 \phi$. One may ask whether there are any other independent Killing vectors.
The answer to this question is in the negative as we now proceed to show.

Before any identifications are made one has the six independent Killing
vectors
 $J_{ab}$ of anti-de-Sitter space. However, after the identifications, not all
 the corresponding vector fields will remain single valued in the quotient
 space.

A necessary and sufficient condition for an adS vector field $\eta$ to
induce a
 well defined vector field on the quotient space is that $\eta$ be invariant
 under the transformation of the identification subgroup,

\bb
(exp 2 \pi \xi)^\ast \eta = \eta
\label{3.40}
\end{equation}
For a Killing vector, this condition becomes

\bb
(exp 2 \pi \xi) \eta (exp 2 \pi \xi)^{-1} = \eta
\label{3.41}
\end{equation}
i.e.
\bb
[exp 2 \pi \xi, \eta] = 0
\label{3.42}
\end{equation}
where $\xi$ and $\eta$ are viewed as $so(2,2)$ matrices.

Now, the matrix $\xi$ can be decomposed as
\bb
\xi = s + n
\label{3.43}
\end{equation}
where (i) $s$ and $n$ commute, (ii) $s$ is semi-simple with real eigenvalues;
and (iii) $n$ is nilpotent (see Appendix A). Accordingly, the semi-simple part
of $(exp 2\pi \xi)$ is $exp 2 \pi s$ and its nilpotent part is $(exp2 \pi s)
[(exp 2 \pi n) - 1]$. Any matrix commuting with $(exp2 \pi \xi)$ must thus
separately commute with $(exp2 \pi s)$ and $(exp2 \pi n)$ (the semi-simple and
nilpotent parts of a matrix can be expressed polynomially in terms of that
matrix). This implies both

\bb
[s ,\eta] = 0
\label{3.44}
\end{equation}
(because the eigenvalues of the matrix $exp2 \pi s$ are real and positive, any
matrix commuting with it must also commute with $log(exp2 \pi s) = 2 \pi s)$
and

\bb
[n, \eta] = 0
\label{3.45}
\end{equation}
(the nilpotent matrix $n$ can be expressed polynomially in
terms of the nilpotent matrix $[(exp2 \pi n) - 1]$ and must thus
commute with $\eta$). It follows from (\ref{3.44}) and
(\ref{3.45}) that $\xi$ and $\eta$ commute,

\bb
[\xi, \ \eta] = 0.
\label{3.46}
\end{equation}
The problem of finding all the Killing vectors of the black
hole solution is thus equivalent to that of finding all the
elements of the Lie algebra $so(2,2)$ that commute with $\xi$.

In order to solve equation (\ref{3.46}) for $\eta$, we observe
that $so(2,2) = so(2,1) \oplus so(2,1)$ and decompose
accordingly $\xi$ into its self-dual and anti-self-dual parts,

\bb
\xi = \xi^+ + \xi^-
\label{3.47}
\end{equation}
Similarly,

\bb
\eta = \eta^+ + \eta^-
\label{3.48}
\end{equation}
The Equation (\ref{3.46}) is equivalent to
\bb
[\xi^+,\eta^+] = 0, \;\;\; [\xi^-,\eta^-] = 0,
\label{3.49}
\end{equation}
because self-dual and anti-self-dual elements automatically commute.
Now, the only elements of $so(2,1)$ that commute with a given non-zero element
of $so(2,1)$ are the multiples of that element. Therefore, since $\xi^+$ and
$\xi^-$ are both non-zero for all values of the black hole parameters we
conclude from (\ref{3.49})

\bb
\eta^+ = \alpha \xi^+, \  \ \eta^- = \beta \xi^- \;\;\;
\alpha, \beta \;\epsilon\;{\bf R}
\label{3.50}
\end{equation}
this shows that the most general Killing vector is a linear combination of
$\partial/\partial t$ and $\partial/\partial \phi$.


\section{Global Structure}
\setcounter{equation} 0

The study of global properties of the 2+1 black hole reveals a
strong coincidence with the 3+1 case. The Penrose diagrams and
maximal extensions are exactly the same as those of a 3+1 black
hole immersed in anti-de Sitter space.

\subsection{Kruskal coordinates}

We follow the analysis of \cite{10}. For the line element
\bb
ds^2  = -(N^{\perp})^2 dt^2 + (N^{\perp})^{-2}dr^2 + r^2(N^{\phi}dt + d\phi)^2
\label{4.1}
\end{equation}
one may introduce a Kruskal coordinate patch around each of the
roots of $(N^{\perp})^2 =0$ to bring the metric to the form

\bb
ds^2= \Omega^2 (du^2-dv^2) +r^2(N^{\phi}dt + d\phi)^2,
\label{4.2}
\ee
where $t=t(u,v)$.

If there is only one root ($J=0$) then the Kruskal coordinates
cover the whole space. When two roots coincide, there are no
Kruskal coordinates \cite{11}.

For definiteness, we start with $r_+$. The Kruskal coordinates
around $r_{+}$ are defined by

Patch $K_{+}$:
\begin{equation} \begin{array}{cl}
r_{-} <r \leq r_{+} & \left\{ \begin{array}{lcr}
U_{+} &=& \left[ \left( \frac{-r+r_{+}}{r+r_{+}} \right)\left(
\frac{r+r_{-}}{r-r_{-}} \right)^{r_{-}/r_{+}} \right]^{1/2}
\sinh a_{+} t  \\
V_{+} &=& \left[ \left( \frac{-r+r_{+}}{r+r_{+}} \right)\left(
\frac{r+r_{-}}{r-r_{-}} \right)^{r_{-}/r_{+}} \right]^{1/2}
\cosh a_{+} t
 \end{array}  \right. \; (a) \\
r_{+} \leq  r < \infty  & \left\{  \begin{array}{lcr}
U_{+} &=& \left[ \left( \frac{r-r_+}{r+r_+} \right)\left(
\frac{r+r_{-}}{r-r_{-}} \right)^{r_{-}/r_{+}} \right]^{1/2}
\cosh a_+ t \\
V_{+} &=& \left[ \left( \frac{r-r_+}{r+r_+} \right)\left(
\frac{r+r_{-}}{r-r_{-}} \right)^{r_{-}/r_{+}} \right]^{1/2}
\sinh a_+ t
\end{array}  \right. \;(b)
\end{array} \label{4.3}
\end{equation}
with
\bb
a_{+} = \frac{r_{+}^2 - r_{-}^2}{l^2 r_{+}},
\label{4.4}
\ee

The angular coordinate (denoted $\phi_{+}$) is chosen on
$K_{+}$ so that the constant of integration
appearing in the solution of (\ref{14}) is fixed to give

\bb
N^{\phi}(r_{+})=0.
\label{4.5}
\ee

The metric takes the form (\ref{4.2}), with the conformal factor

\begin{equation}
\Omega^2 (r)= \frac{(r^2-r_{-}^2)(r+r_{+})^2}{a_{+}^2 r^2 l^2} \left(
\frac{r-r_{-}}{r+r_{-}} \right)^{r_{-}/r_{+}} \;\;\;  r_{-}
< r <\infty .
\label{4.6}
\end{equation}

With the choice of $\phi$ leading to (\ref{4.5}), the term
$N^{\phi} dt$ in (\ref{4.2}) remains regular at $r_+$.

Similarly, around $r_{-}$, one defines

Patch $K_{-}$:

\begin{equation} \begin{array}{cl}
0 <r \leq r_{-} & \left\{ \begin{array}{lcr}
U_{-} &=& \left[ \left( \frac{-r+r_{-}}{r+r_{-}} \right)\left(
\frac{r+r_{+}}{-r+r_{+}} \right)^{r_{+}/r_{-}} \right]^{1/2}
\cosh a_- t  \\
V_{-} &=& \left[ \left( \frac{-r+r_{-}}{r+r_{-}} \right)\left(
\frac{r+r_{+}}{-r+r_{+}} \right)^{r_{+}/r_{-}} \right]^{1/2}
\sinh a_- t
 \end{array}  \right. \; (a) \\
r_{-} \leq r \leq r_{+} & \left\{  \begin{array}{lcr}
U_{-} &=& \left[ \left( \frac{r-r_-}{r+r_-} \right)\left(
\frac{r+r_{+}}{-r+r_{+}} \right)^{r_{+}/r_{-}} \right]^{1/2}
\sinh a_- t \\
V_{-} &=& \left[ \left( \frac{r-r_-}{r+r_-} \right)\left(
\frac{r+r_{+}}{-r+r_{+}} \right)^{r_{+}/r_{-}} \right]^{1/2}
\cosh a_{-} t
\end{array}  \right. \;(b)
\end{array} \label{4.7}
\end{equation}
with

\bb
a_{-} = \frac{r_{-}^2 - r_{+}^2}{l^2 r_{-}}.
\label{4.8}
\ee
This time, one chooses the angular coordinate $\phi_{-}$ so that
$N^{\phi}(r_{-}) = 0$. The metric takes the form (\ref{4.2}) with

\begin{equation}
\Omega^2 (r)= \frac{(r_{+}^2-r^2)(r+r_{-})^2}{a_{-}^2 r^2 l^2} \left(
\frac{r_{+}-r}{r_{+}+r} \right)^{r_{+}/r_{-}} \;\;\; 0< r < r_{+}.
\label{4.9}
\end{equation}

The overlap  of the patches $K_{+}$ and $K_{-}$
($r_{-}<r<r_{+}$) will be called $K$. Just as in the 3+1 case
one may maximally extend the geometry by glueing together an
infinite number of copies of patches $K_{+}$, $K_{-}$. We will
not illustrate graphically that extension in terms of Kruskal
coordinates, but will rather go to the more economical Penrose
diagrams.


\subsection{Penrose diagrams ($r_+ \neq r_-$)}

The Penrose diagrams are obtained by the usual change of coordinates

\begin{equation}
U+V = \tan\left(\frac{p+q}{2}\right) \;\;\;\; U-V =
\tan\left(\frac{p-q}{2}\right).
\label{4.10}
\end{equation}
We define the inverse transformation by taking the usual determination of the
 inverse tangent, namely the one that lies between $- \pi/2$ and $+ \pi/2$.

Consider first the case $J=0$. From (\ref{4.10}) and (\ref{4.3})
(with $r_-=0$) it is easy to prove that, (i) $r=\infty$ is mapped to
the lines $p=\pm \m\pi$, (ii) the singularity $r=0$ is mapped to
the lines $q=\pm \m \pi$ and (iii) the horizon is mapped to $p=\pm
q$. The Kruskal and Penrose diagrams associated with this
geometry are shown in Fig.3.

Next consider the case of the rotating black hole. By making the
change of coordinates (\ref{4.10}) in the two patches defined in
Sec.(4.1) we find one Penrose diagram for each patch.  These are
shown in Figs. (4a, b).

The regions shown as $K$ in parts (a) and (b) of Fig.4 are to be
identified because they are the overlap. Now, the original black
hole coordinates covered $K$ and one region {\bf III} in (4.a),
and $K$ and one region {\bf I} in (4.b). However, one wants to
obtain a ``maximal causal extension" (i.e., a maximal extension
without closed timelike curves). To this effect one must first
include the other two regions in each diagram and then glue
together an infinite sequence of them, as shown in Fig.(4.c).

\subsection{Extreme cases $M=0$ and $M=|J|/l$}

\subsubsection{{\bf $M=0$}}

The metric is
\bb
ds^2 = -(r/l)^2 dt^2 + (r/l)^{-2} dr^2 + r^2 d\phi^2.
\label{4.11}
\ee
Defining the null dimensionless coordinates
\bb
u=\frac{t}{l}-\frac{l}{r}, \;\;\; v=-\frac{t}{l}-\frac{l}{r}
\label{4.12}
\ee
we find

\bb
ds^2= r^2dudv + r^2d\phi^2.
\label{4.13}
\ee
and pass directly to Penrose coordinates by

\bb
U=\tan\m(p+q), \;\;\; V=\tan\m(p-q).
\label{4.14}
\ee
The relation between the radial coordinate $r$ and $p,q$ is

\bb
-r=l\frac{\cos p + \cos q}{\sin p},
\label{4.15}
\ee
and the metric takes the form

\bb
ds^2 = l^2\frac{dp^2 - dq^2}{\sin^2 p} + r^2 d\phi^2.
\label{4.16}
\ee
 From (\ref{4.15}) it is easy to show that the origin is mapped
to the segment of the lines $p=\pi \pm q$ running from $p=0$ to
$p=\pi$ while spacelike infinity is mapped to the segment of the
$p=\pi$ line that closes the triangle shown in Fig.(5a).


\subsubsection{{\bf $M=|J|/l$}}

The metric is
\bb
ds^2 = -\frac{(r^2-r_+^2)^2}{r^2 l^2}dt^2 +
\frac{r^2 l^2}{(r^2-r_+^2)^2}dr^2 + r^2(N^{\phi}dt + d\phi)^2
\label{4.17}
\ee
where $r=r_+=l (M/2)^{1/2}$ is the horizon. Introducing the
null coordinates $U=t+r^{*}$ and $V=-t+r^{*}$ where $r^{*}$ is
the tortoise coordinate

\bb
r^{*}=\int \frac{dr}{(N^{\perp})^2}= \frac{-rl^2}{2(r^2-r_+^2)} +
\frac{l^2}{4r_+}\ln\left|\frac{r-r_+}{r+r_+} \right|
\label{4.18}
\ee
and defining the Penrose coordinates $p,\, q$ as in (\ref{4.14}) we
obtain the line element

\bb
ds^2 = \frac{4(N^{\perp})^2 l^2(dp^2-dq^2)}{(\cos p + \cos q)^2}
+ r^2 (N^{\phi} dt + d\phi)^2.
\label{4.19}
\ee

{}From
\bb
\frac{\sin p}{\cos q + \cos p} =  \frac{-rl}{2(r^2-r_+^2)} +
\frac{l}{4r_+}\ln\left|\frac{r-r_+}{r+r_+} \right|,
\label{4.20}
\ee
one sees that the lines $r=r_{+}$ are at $\pm
45^{\circ}$, whereas $r=0$ is at $p=(k \pi)^{+}$ and
$r=\infty$ at $p=(k \pi)^{-}$. [By $p = (k \pi)^{+}$, we
mean that $r \rightarrow 0$ as $p \rightarrow k \pi$ from
value greater than $k \pi$, and similarly, $r \rightarrow
\infty$ as $p \rightarrow k \pi$ from
values smaller than $k \pi$].  If we take for $p$ the usual
determination of the arc tangent in (\ref{4.14}), so that the
region $0<r<r_{+}$ is mapped on the triangle bounded by $p=0 \;
(r=0)$ and $p = q = \pi, \; p-q = \pi$, then we must take in the
region $r>r_{+}$ \underline{a different determination}. Indeed,
one must glue the triangle corresponding to $r > r_{+}$ to the
triangle corresponding to $0<r<r_{+}$ along the sides $r=r_{+}$
at 45$^{\circ}$, and not along the vertical sides (which are $r=
\infty$ in the region $r>r_+$ and $r=0$ in the region $r<r_{+})$. For
instance, one could map $r>r_{+}$ into the triangle bounded by
$p+q=\pi, p-q= - \pi$ and $p=\pi$. Once this is done, one can go
safely across $r=r_{+}$ because the zero of $N^{\perp}$ in
(\ref{4.19}) is cancelled by the zero in the denominator. To
achieve the maximal extension one then needs to include an
infinite sequence of triangles as shown in Fig.(5.b) (the
original black hole geometry just included two adjacent
triangles).


\section{Instability of metric regularity at $r^2=0$. \newline
Chronology Protection}

 The point of view taken in this article is that the region $r^2<0$  must be
cut out from the spacetime because it contains
closed timelike lines (see Fig.6 for a Penrose diagram that includes the
forbidden region). This is a consistent point of view and leads to a close
analogy with the black hole in  3+1 dimensions. There is, however, a
compelling additional argument for considering the spacetime as ending at
$r=0$. It is the fact that the introduction of matter produces a curvature
singularity at $r=0$. This can be easily seen in simple examples and we
believe it to be a general feature (with the possible exception of very
``fine-tuned" couplings).
  The first example is the collapse of a cloud of dust with $J=0$ \cite{12}.
One can then verify that the matter will reach infinite density at r=0.
In this case only the part of the surface $r=0$ that intersects the history
of the dust becomes singular. This is due to the fact that the dust ``probes"
only part of the spacetime. However, in the case of a field- such as the
electromagnetic field - which is our second example - all the spacetime is
 probed.
   As it was indicated in\cite{1}, the introduction of a Maxwell field that
depends
only on the radial coordinate yields an electromagnetic field for which
the gauge invariant scalar $F_{\mu \nu} F^{\mu \nu}$ is proportional to
$r^{-2}$ and thus is singular at all points on the surface $r=0.$

Therefore, in view of the curvature singularities that are
brought in by matter couplings, it seems not only reasonable,
but also compulsory, to exclude the region $r^{2} < 0$ from the
spacetime.

  The collapsing dust is also interesting in that it may be regarded as a
mechanism for producing, without effort, closed timelike lines from a
perfectly
reasonable initial condition ( with the help of a negative cosmological
constant though!). However, one sees, first of all, that the closed timelike
lines are hidden behind the horizon at $r=r_+ >0$ (Fig. 7). But,
moreover, if - say - an electromagnetic field is brought in, a barrier of
 infinite
curvature is introduced at $r=0$. This makes the closed timelike lines not
reachable from $r^2 > 0$.  In this sense we see that
``chronology is protected" \cite{13} in the 2+1 black-hole.

\newpage
{\bf Acknowledgements}\\

Informative discussions with Steven Carlip, Frank Wilczek,
and Edward Witten are gratefully acknowledged. M. B. holds
a Fundaci\'on Andes Fellowship and M.H. gratefully
acknowledges the hospitality of the Institute for Advanced
Study where the research reported in this paper was
partially carried out.  This work was supported in part by
grants 0862/91 and 0867/91 of FONDECYT (Chile), grant PG/082/92
of Departamento Postgrado y Post\'{\i}tulo, Universidad de Chile,
by research funds from F.N.R.S (Belgium), by a European Communities research
contract, and by
institutional support provided by SAREC (Sweden) and
Empresas Copec (Chile) to the Centro de Estudios
Cient\'{\i}ficos de Santiago.

\newpage
\noindent {\Large {\bf Appendix A.~~
One Parameter Subgroups of $SO(2,2)$}}


\renewcommand{\theequation}{A.\arabic{equation}}
\setcounter{equation} 0
\noindent {\large {\bf A.1 ~~Description of the problem}}

The purpose of this Appendix is to provide a complete
classification of the inequivalent one-parameter subgroups of
$SO(2,2)$. Two one-parameter subgroups $\{g(t)\}$ and
$\{h(t)\}$,  $t\epsilon {\bf R}$, are said to be equivalent if
and only if they are conjugate in $SO(2,2)$, i.e.,

\bb
g(t)= k^{-1}h(t)k,\;\;\; k\epsilon SO(2,2)
\label{a.1}
\ee

By an $SO(2,2)$ rotation of the coordinate axes in ${\bf R}^4$,
one can then map $g(t)$ on $h(t)$. Since one-parameter subgroups are
obtained by exponentiating infinitesimal transformations, the
task at hand amounts to classifying the elements of the Lie
algebra $so(2,2)$ up to conjugation.

Now, the elements of $so(2,2)$ are described by antisymmetric
tensors $\omega_{ab} =-\omega_{ba}$ in ${\bf R}^4$. If one
conjugates the infinitesimal transformation $R^a\;_b =
\delta^a\;_b +\varepsilon \omega^a\;_b$ by $k \; \epsilon
SO(2,2)$, ($k^T \eta k = \eta,\;\; \eta= \mbox{diag}(--++)$),
one finds that the antisymmetric matrix $\omega \equiv
(\omega_{ab})$ transforms as

\bb
\omega\rightarrow  \omega'= k^T \omega k,\;\;k \epsilon SO(2,2)
\label{a.2}
\ee

Hence we have to classify antisymmetric tensors under the
equivalence relation (\ref{a.2}).
\vspace{1cm}


\noindent {\large {\bf A.2 ~~Strategy}}

Any linear operator $M$ can be uniquely decomposed as the sum of
a semi-simple (diagonalizable over the complex numbers) linear
operator $S$ and a nilpotent operator $N$ that commute,

\bb
M=S+N,
\label{a.3}
\ee
\bb
[S,N]=0
\label{a.4}
\ee
with

\bb
N^q=0 \;\;\; \mbox{for some q}
\label{a.5}
\ee
and

\bb
S= L^{-1} (\mbox{diagonal matrix}) L,\;\; \mbox{for some}\; L
\label{a.6}
\ee
(Jordan - Chevalley decomposition of $M$).

The eigenvalues of $S$ coincide with those of $M$ and provide an
intrinsic characterization of $S$. When the eigenvalues of $S$
are non-degenerate, the nilpotent operator $N$ is identically
zero and $M$ is thus completely characterized (up to similarity)
by its eigenvalues.  However, if some eigenvalues are repeated,
$N$ may be non-zero and $M$ cannot be reconstructed from the
knowledge of its eigenvalues: one needs also information about
its nilpotent part (the dimensions of the irreducible invariant
subspaces).

We shall construct the sought-for invariant classification of
elements of $so(2,2)$ by means of the Jordan - Chevalley
decomposition of the operator $\omega^a\,_b$.Since
$\eta^{ab}\neq \delta_{ab}$ for $SO(2,2)$, the operator
$i\omega^a\,_b$ is, in general, not hermitian. Accordingly, it
may possess a non-trivial nilpotent part when its eigenvalues
are degenerate. The classification of the possible $\omega^a\,_b$
is analogous to the invariant classification of the
electromagnetic field in Minkowski space and is also reminiscent
of the Petrov classification of the Weyl tensor in General
Relativity.

Because the matrix $\omega_{ab}$ is real and antisymmetric,
there are restrictions on its eigenvalues. These constraints are
contained in the following elementary Lemmas.

{\bf Lemma 1}: If $\lambda$ is an eigenvalue of $\omega_{ab}$, then
$-\lambda$ is also an eigenvalue of $\omega_{ab}$.\\

{\bf Proof}: From

\bb
(\omega_{ab} -\lambda \eta_{ab})l^{b} =0
\label{a.7}
\ee
one infers the characteristic equation

\bb
\mbox{det}(\omega -\lambda \eta)=0
\label{a.8}
\ee

But then $ 0=\mbox{det}(\omega-\lambda \eta)^T
=\mbox{det}(-\omega-\lambda
\eta)=\mbox{det}(\omega +\lambda \eta)$, i.e., $-\lambda$ is
also a root of the characteristic equation.$\Box$

{\bf Lemma 2}: If $\lambda$ is an eigenvalue, then $\lambda$* is also
an eigenvalue.\\

{\bf Proof}: This is a consequence of the reality of $\omega_{ab}$,
which implies that the characteristic equation (\ref{a.8}) has
real coefficients.$\Box$

\vspace{1cm}
\noindent {\bf A.2.1 ~~Types of eigenvalues}

It follows from these theorems that the four eigenvalues of
$\omega$ are of the following four possible types:

\begin{enumerate}
\item $ \lambda,\;\; -\lambda,\;\; \lambda^{*},\;\;
-\lambda^{*},\;\;\;\; \lambda=a +ib,\;\;\;\; a\neq 0
\neq b$

\item $\lambda_1 = \lambda_1^{*},\;\; -\lambda_1,\;\; \lambda_2
= \lambda_2^{*},\;\; -\lambda_2,\;\; \;\;(\mbox{$\lambda_1$  and
$\lambda_2$  real})$

\item $ \lambda_1,\;\; -\lambda_1 = \lambda_1^{*},\;\; \lambda_2,\;\;
-\lambda_2 =  \lambda^{*} ,\;\;\;\;(\mbox{$\lambda_1$ and $\lambda_2$
imaginary})$

\item $ \lambda_1 =\lambda_1^{*},\;\;  -\lambda_1,\;\; \lambda_2,\;\;
-\lambda_2 =\lambda_2^{*},\;\;\;\; (\mbox{$\lambda_1$ real,
$\lambda_2$ imaginary})$
\end{enumerate}

In each case, the eigenvalues involve only two independent real
numbers, whose knowledge is equivalent to knowing the two
Casimir invariants.

\bb
I_1 = \omega^{ab}\omega_{ab}, \;\;\; I_2 = \m \epsilon^{abcd}
\omega_{ab} \omega_{cd}
\label{a.9}
\ee
[If one replaces $SO(2,2)$ by $SO(4)$, $i\omega^a\,_b$ is
hermitian and therefore diagonalizable. Hence there is no
nilpotent part and $i\omega^a\,_b$ is completely
characterized by its eigenvalues and thus by $I_1$ and $I_2$.]

Multiple roots can occur only in the following circumstances:

\begin{itemize}
\item Cases (2) and (3), when $\lambda_1= \lambda_2$ (or $-\lambda_2$). If
$\lambda_1 \neq 0$, then $\lambda_1$ and $-\lambda_1$ are
distinct roots. If $\lambda_1=0$, then 0 is a quadruple root; or

\item Cases (2),(3) or (4), when one of the roots vanishes.
\end{itemize}
\vspace{1cm}


\noindent {\bf A.2.2 ~~Types of antisymmetric tensors}

For simple roots, one can give a unique canonical form to which
any matrix $\omega_{ab}$ with a given set of eigenvalues can
be brought to by an $SO(2,2)$ transformation. This is the form
of $\omega_{ab}$ in the basis where $\omega^a\,_b$ is
diagonal.  In the presence of multiple roots, there are
inequivalent canonical forms because $\omega^a\,_b$ may contain a
non-trivial nilpotent part $N$. But for each possible type of
$N$, there is a unique canonical form.  These canonical forms
are all derived in the next subsections.

We shall say that the matrix $\omega^{ab}$ is of type $k$ if its
nilpotent part is of order $k$, $N^k =0$. The types {\bf I} and
{\bf II} can be further classified according to the reality
properties of the roots. We thus define:
\\
\noindent{{\bf  Type I}  $(N=0)$}

{\bf I$_a$}: 4  complex roots $\lambda,\; -\lambda,\;
\lambda^*,\; -\lambda^*\; (\lambda \neq \pm \lambda^*)$.

{\bf I$_b$}: 4 real roots $\lambda_1,\; -\lambda_1,\;
\lambda_2, -\lambda_2$.

{\bf I$_c$}: 4 imaginary roots $\lambda_1,\; -\lambda_1,\;
\lambda_2,\; -\lambda_2$.

{\bf I$_d$}: 2 real ($\lambda_1$ and $-\lambda_1$), and two
imaginary roots ($\lambda_2$ and $-\lambda_2$).

\noindent{{\bf Type II} $(N \not= 0, N^2 = 0)$}

{\bf II$_a$}: 2 real double roots, $\lambda$ and
$-\lambda$.

{\bf II$_b$}: 2 imaginary double roots, $\lambda$ and
$-\lambda$.

{\bf II$_c$}: 1 double root (0) and 2 simple roots
($\lambda$ and $-\lambda$, with $\lambda$ real or imaginary.)

\noindent{{\bf Type III} $(N^2 \not= 0, N^3 =0)$: one quadruple root,
 zero.}

\noindent{{\bf Type IV} $(N^3 \not= 0, N^4 =0)$: one quadruple root, zero.}
\\
We shall write in all cases

\bb
\lambda = a+ib
\label{a.9.1}
\ee

We close this section by proving the following useful Lemma.

{\bf Lemma 3}: Let $v^a$ and $u^a$ be eigenvectors of $\omega
^a\;_b$ with respective eigenvalues $\lambda$ and $\mu$,

\bb
\omega^a\,_b v^b = \lambda v^a, \;\;\; \omega ^a\,_b u^b = \mu u^a.
\label{a.10}
\ee
Then $v_a u^a = 0$ unless $\lambda + \mu =0$. In particular, if
$\lambda \neq 0$, then $v^a$ is a null vector.

{\bf Proof}: One has $u_a\omega^a\,_b v^b = \lambda u_a v^a =
-\mu u^a v_a$, and thus $(\lambda+\mu)  u^a v_a=0$. $\Box$

We now proceed to the explicit determination of the canonical forms.

\vspace{1cm}


\noindent {\large {\bf A.3~~ Type I$_a$}}

One has by definition of type {\bf I}$_a$,

\begin{eqn}
\omega _{ab} l^b &=& \lambda l_a
\aum \label{a.11.a} \\
\omega_{ab} m^b &=& -\lambda m_b
\aum \label{a.11.b} \\
\omega_{ab} l^{*b} &=& \lambda^* l^*_a
\aum \label{a.11.c} \\
\omega_{ab} m^{*b} &=& -\lambda^* m^*_a
\aum \label{a.11.d}
\end{eqn}

\noindent where the eigenvectors $l^a,\;l^{*a},\;m^a,\;m^{a*}$ are complex
and linearly independent. The only scalar products that can be
different from zero are $l^a m_a$ and $l^{a*}m_a^*$. They cannot
vanish since the metric would then be degenerate. By scaling
$m_a$ if necessary one can assume $l^a m_a =1$. One then has
also $l^{a*} m_a^* =1$. The metric is given by

\bb
\eta _{ab} = l_a m_b + l_a^* m_b^* + [a\leftrightarrow b]
\label{a.12}
\ee
since
$$(\eta_{ab} - l_a m_b-l_a^* m_b^* - [a\leftrightarrow
b])\,u^b$$
is zero whenever $u^a$ equals $l^a,\, m^a,\, l^{a*},\,
m^{b*}$. The tensor $\omega^{ab}$ is given by

\bb
\omega^{ab} = \lambda(l_a m_b -l_b m_a) + \lambda^* (l_a^* m_b^* - l_b^*
m_a^*)
\label{a.13}
\ee
because this reproduces (\ref{a.11.a})-(\ref{a.11.d}).

Our goal is to achieve a canonical expression for $\omega^a\,_b$
over the real numbers. Therefore we decompose the vectors $l_a$ and
$m_a$ into their real and imaginary components

\bb
l_a = u_a +iv_a, \;\; m_a =n_a +iq_a,
\label{a.14}
\ee
(the transformation $l_a,\; m_a,\; l_a^*,\;m_a^* \rightarrow
u_a,\; v_a,\; n_a,\; q_a$ is invertible and so, the vectors
$u_a,\; v_a,\; n_a$, and $q_a$ form a basis). This gives

\br
\eta_{ab} &=& 2(u_a n_b -v_a q_b) + [a\leftrightarrow b] \\
\omega_{ab} &=& 2a(u_an_b -v_aq_b) -2b(u_a q_b +v_a n_b) -
[a\leftrightarrow b]
\er
In the orthonormal basis  where the vectors $u_a, \;
v_a,\; n_a,\; q_a$, have components $u_a =(0,\m,\m,0),\;\;
n_a=(0,-\m,\m,0),\;\; v_a=(\m,0,0,\m),\;\;$ and $q_a=(\m,0,0,-\m)$,
$\omega_{ab}$ take the form

\begin{eqn}
\omega_{ab} &=& \left(\begin{array}{rrrr}

               0 &  b  &  0  &  a \\
              -b &  0  &  a  &  0 \\
               0 & -a  &  0  &  b \\
              -a &  0  & -b  &  0
\end{array} \right)
\label{a.17}
\end{eqn}
Eq. (\ref{a.17}) is the canonical form of an antisymmetric
tensor of type {\bf I}$_a$. The Casimir invariants are found from
(\ref{a.9}) to be

\begin{eqn}
I_1 &=& 4 (b^2 - a^2)
\aum \label{a.18.a}) \\
I_2 &=& 4(b^2+a^2)
\aum \label{a.18.b})
\end{eqn}

\vspace{1cm}

\noindent {\large {\bf A.4 ~~Type {\bf I}$_b$}}

One has, by definition of type I$_b$,

\begin{eqn}
\omega_{ab} l^b &=& \lambda_1 l_a,\;\; \omega_{ab} m^b =
-\lambda_1 m_a
\aum \label{a.19.a}\\
\omega_{ab} n^b &=& \lambda_2 n_a,\;\; \omega_{ab} u^b =
-\lambda_2 u_a
\aum \label{a.19.b}
\end{eqn}

The vectors $l^a,\; m^a,\; n^a,\;$ and $u^a$ are real and
linearly independent, and the non-vanishing scalar products are
$l\cdot m$ and $n\cdot u$. Straightforward steps yield then, in
an orthonormal basis, the canonical form

\begin{eqn}
\omega_{ab} &=& \left(
              \begin{array}{rrrr}
               0 &  0  &  0  & -\lambda_2 \\
               0 &  0  & -\lambda_1  &  0 \\
               0 &  \lambda_1  &  0  &  0 \\
	           \lambda_2 &  0  &  0  &  0
\end{array} \right)
\label{a.20}
\end{eqn}

The Casimir invariants are given by

\begin{eqn}
I_1 &=& -2(\lambda_1^2 +\lambda_2^2),
\aum \label{a.21.a} \\
I_2 &=& 4\lambda_1 \lambda_2.
\aum \label{a.21.b}
\end{eqn}

\vspace{1cm}


\noindent {\large {\bf A.5 ~~Type I$_c$}}

One has, by definition of type {\bf I}$_c$

\begin{eqn}
\omega_{ab} l^b &=& ib_1 l_a,\;\; \omega_{ab} l^{b*} =
-ib_1 l_a^*
\aum \label{a.22.a} \\
\omega_{ab} m^b &=& ib_1 m_a,\;\; \omega_{ab} m^{b*} =
-ib_1 m_a^*
\aum \label{a.22.b}
\end{eqn}

The only non-vanishing scalar products are $l_a l^{a*}$ and
$m_a m^{a*}$. One can rescale $l_a$ and $m_a$ so that $l\cdot
l^* =\pm 1, \; m\cdot m^* =\mp 1$. If $l\cdot l^* = 1$, then
$m\cdot m^* = -1$ and vice versa. [Through
$l^a=\frac{1}{\sqrt{2}}(u^a +iv^a)$, one associates to a vector
$l^a$ obeying $l_a l^{a*}=1$, two real vectors $u^a$, $v^a$ ,
such that $u_a u^a =\; 1\; = v_a v^a, \;\; u_a v^a = 0$. So, if
$l_a l^{a*}=1$, one must have  $m_a m^{a*} =-1$ in order to
agree with the signature $(- - + +)$ of the metric.]

One obtains the final canonical form

\begin{eqn}
\omega_{ab} &=& \left(
              \begin{array}{rrrr}

               0    &  b_1  &   0    &  0   \\
              -b_1  &  0    &   0    &  0   \\
               0    &  0    &   0    &  b_2 \\
	           0    &  0    &  -b_2  &  0
\end{array} \right)
\label{a.23}
\end{eqn}
for $\omega_{ab}$ in a real orthonormal basis.

The Casimir invariants are found to be

\begin{eqn}
I_1 &=& 2(b_1^2  +b_2^2),
\aum \label{a.24.a} \\
I_2 &=& 4b_1 b_2.
\aum \label{a.24.b}
\end{eqn}

\vspace{1cm}

\noindent {\large {\bf A.6 ~~Type I$_d$}}

Type {\bf I}$_d$ does not exist. Indeed, the real eigenvalue
brings a block of signature $(+ -)$, while the imaginary
eigenvalue brings a block of signature $(+ +)$ or $(- -)$. This
is inconsistent with signature $(- - + +)$.

\vspace{1cm}

\noindent {\large {\bf A.7 ~~Role of the Casimir invariants for type {\bf I}}}

If one compares (\ref{a.18.a}), (\ref{a.18.b}),
(\ref{a.21.a}), (\ref{a.21.b}) and (\ref{a.24.a}), (\ref{a.24.b}),
one sees that the Casimir invariants completely characterize the
matrices $\omega_{ab}$ of type {\bf I}. If I$_1 \pm$I$_2$ are
both positive, the type is type {\bf I}$_c$. If I$_1 \pm$I$_2$
are both negative, the type is type {\bf I}$_b$. Otherwise, the
type is I$_a$. Furthermore, the eigenvalues can be reconstructed
from I$_1$ and I$_2$. The roots are degenerate when I$_1 +$I$_2$
or I$_1 -$I$_2$ vanish.  It is easy to see that I$_1 \pm$I$_2$
are the Casimir invariants of the two algebras $so(2,1)$ contained in
$so(2,2,)=so(2,1)\oplus so(2,1)$. The self-dual and
anti-self-dual (real) matrices $\omega_{ab}^{\pm} =\omega_{ab}
\pm \m \epsilon_{ab}\,^{cd}
\omega _{cd}$ define irreducible representations of $so(2,2)$
($\omega_{ab}^{+}$ transforms as a vector under the first
$so(2,1)$, while $\omega_{ab}^{-}$ transforms as a vector under
the second.) One has $2I_1 = \omega_{ab}^{+}
\omega^{+ab}$ and $2I_2= \omega_{ab}^{-} \omega^{-ab}$. There is
however, no particular advantage in working with the self-dual and
anti-self-dual components of $\omega_{ab}$ in the subsequent
discussion. For that reason, we shall not perform the split.

\vspace{1cm}

\noindent {\large {\bf A.8 ~~Type II$_a$}}

By definition of type {\bf II}$_a$, there are two doubly degenerate,
non-zero, real eigenvalues $\lambda$ and $-\lambda$. Each
eigenvalue has at least one eigenvector, thus one can find $l^a$
and $m^a$ such that

\begin{eqn}
\omega_{ab} l^b &=& \lambda l_a
\aum \label{a.25.a} \\
\omega_{ab} m^b &=& -\lambda m_a
\aum \label{a.25.b}
\end{eqn}

Within each invariant subspace we can introduce an additional
vector to complete $l,\;m$ to a basis. Since  $\omega^a\;_b$
has a nilpotent part, at least one of the additional vectors
will not be an eigenvector. We can thus write, without loss of
generality,

\begin{eqn}
\omega_{ab}u^b &=& \lambda u_a + l_a
\aum \label{a.26.a}\\
\omega_{ab}s^b &=& -\lambda s_a +\alpha m_a
\aum \label{a.26.b}
\end{eqn}

It follows from (\ref{a.25.a}), (\ref{a.25.b}) and
(\ref{a.26.a}), (\ref{a.26.b}) that $l\cdot l=\; l\cdot m=\;
l\cdot u=0$. Hence, since the metric is non-degenerate we must
have $l\cdot s \neq 0$. This implies in turn that $\alpha$ must
be different from zero since (\ref{a.26.a}), (\ref{a.26.b})
gives $l\cdot s+\alpha m\cdot u=0$. By a rescaling of $m$ we can
set $\alpha =1$, so one has

\bb
\omega_{ab} s^b = -\lambda s_a +m_a
\label{a.27}
\ee

The remaining scalar products are evaluated as follows. First,
one can take $u^a s_a=0$ since one can redefine $u^a \rightarrow
u^a + \rho l^a$ without changing any of the previous relations.
Second, by multiplying (\ref{a.26.a}) with $u^a$, one gets,
using $u^al_a =0$, that $u^au_a=0$. One then finds
from (\ref{a.27}) $u^a m_a =-1$ as the only remaining
non-vanishing scalar product.

The metric and antisymmetric tensor $\omega_{ab}$ read

\begin{eqn}
\eta_{ab} &=& l_as_b - m_a u_b + [a\leftrightarrow b]
\aum \label{a.28.a} \\
\omega_{ab} &=& \lambda(l_as_b - u_a m_b) - l_a m_b - [a\leftrightarrow b]
\aum \label{a.28.b}
\end{eqn}

In a suitable orthonormal frame, this gives

\begin{eqn}
\omega_{ab} &=& \left(
              \begin{array}{rrrr}
               0   &  1  &  1   & \lambda   \\
              -1   &  0  & \lambda  & 1   \\
              -1   & -\lambda   &  0  & 1 \\
	          -\lambda & -1 & -1 &  0
\end{array} \right)
\label{a.29}
\end{eqn}

When $\lambda \neq 0$, a simpler, equivalent canonical form, can be achieved
by replacing $m_a$ by $m'_a +l_a /2\lambda$ and $s_a$ by $s'_a +
u_a /2\lambda$.  This leaves $\eta_{ab}$ unchanged

\bb
\eta_{ab} = l_a s'_b- m'_a u_b + [a\leftrightarrow b],
\label{a.30}
\ee
but modifies $\omega_{ab}$ to

\bb
\omega_{ab}=\lambda(l_a s'_b-u_a m'_b) + l_a (u_b-m'_b) - [a
\leftrightarrow b],
\label{a.31}
\ee
which, in an appropriate orthonormal frame, yields

\begin{eqn}
\omega_{ab} &=& \left(
              \begin{array}{rrrr}
               0   &  0  &  0  & \lambda   \\
               0   &  0  & \lambda    & 1   \\
               0   & -\lambda   &  0  & 1 \\
	      -\lambda & -1         & -1  &  0
\end{array} \right)
\label{a.32}
\end{eqn}
\\
The forms (\ref{a.29}) and (\ref{a.32}) are not equivalent when
$\lambda=0$. It is only (\ref{a.29}) that is if type II$_a$ in
that case, since (\ref{a.32})
 with $\lambda = 0$ possesses a non trivial nilpotent part of order 3 and is
 thus of type III. The Casimir invariants
are found to be

\begin{eqn}
I_1 &=& -4\lambda^2,
\aum \label{a.33.a} \\
I_2 &=& 4\lambda^2.
\aum \label{a.33.b}
\end{eqn}
\\
i.e., they are exactly the same as those of (\ref{a.21.a}),
(\ref{a.21.b}) with $\lambda_1=\lambda_2$. However, the
canonical forms (\ref{a.29}) or (\ref{a.32}) are not equivalent
to (\ref{a.20}) with $\lambda_1=\lambda_2$ since they possess a
non trivial nilpotent part , while (\ref{a.20}) does not for any
value of $\lambda_1$, $\lambda_2$.

\vspace{1cm}

\noindent {\large {\bf A.9 ~~ Types II$_b$ and II$_c$}}

The analysis of type {\bf II}$_b$ proceeds as for type {\bf II}$_a$. We only
quote the final canonical form in an orthonormal basis

\begin{eqn}
\omega_{ab} &=& \left(
				\begin{array}{cccc}
				  0     &    b-1    & -1    & 0 \\
				-b+1    &    0      &  0    & 1 \\
				  1     &    0      &  0    & b+1 \\
				  0     &   -1      & -b-1  & 0
\end{array} \right)
\label{a.34}
\end{eqn}
and the Casimir invariants

\bb
I_1 = 4b^2,\;\;\; I_2=4b^2.
\label{a.35}
\ee

Type {\bf II}$_c$ is incompatible with a non-degenerate metric
and so it does not exist. Indeed, the equations
$\omega_{ab}l^b=0$, $\omega_{ab}m^b=l_a$ (0 is a double root and
$\omega^a_b$ is a non trivial nilpotent matrix in the
corresponding invariant eigenspace), together with
$\omega_{ab}u^b =\lambda u_a$, $\omega_{ab}v^b =-\lambda v_a$
imply $l\cdot l = l\cdot \omega\cdot m=-(l\omega)\cdot m=0$,
$l\cdot m=\omega\cdot m=0$, $l\cdot u = \lambda^{-1} l\cdot
\omega\cdot u=0$, $l\cdot v =-\lambda^{-1} l \cdot v =0$.
So $l^a$ would be a non zero vector orthogonal to any vector and
the metric would be degenerate.

\vspace{1cm}

\noindent {\large {\bf A.10 ~~Types III and IV}}

In type {\bf III}, zero is a quadruple root of the characteristic
equation. Since $\omega^a\;_b$ is nilpotent of order 3, one can
find a basis such that

\begin{eqn}
\omega_{ab}l^b = 0
\aum \label{a.36.a}\\
\omega_{ab}m^b = 0, \;\;\; \omega _{ab}u^b=m_a, \;\;\;  \omega_{ab}t^b=u_a.
\aum \label{a.36.b}
\end{eqn}

The scalar product of $l^a$ with $u_a$ vanishes from
(\ref{a.36.b}). Similarly, $m\cdot m= m\cdot u=0$. Hence $m \cdot
t$ cannot vanish, say $m \cdot t = \pm 1$. Then, by a
redefinition of $l^a$, $l^a \rightarrow l^a + \rho m^a$, one can
assume $l\cdot t =0$. It follows that $l\cdot l\neq 0$ since
otherwise the metric would be degenerate. We set $l\cdot l =
-\varepsilon$, $\varepsilon =\pm 1$. By making appropriate
redefinitions of $t^a$ if necessary and using the fact that the
metric is of signature $(--++)$, one finally obtains

\begin{eqn}
\eta_{ab} &=& \varepsilon( -l_a l_b - m_a t_b -t_b m_a +u_a u_b)
\aum \label{a.37.a} \\
\omega_{ab} &=& \varepsilon(m_a u_b - u_a m_b)
\aum \label{a.37.b}
\end{eqn}
This yields in an appropriate orthonormal basis

Type {\bf III}$^+$ ($\varepsilon=+1)$.

\begin{eqn}
\omega_{ab} &=& \left(
\begin{array}{cccc}
  0     &    0         &  0   & 0 \\
  0     &    0         &  0   & 1 \\
  0     &    0         &  0   & 1 \\
  0     &   -1         & -1   & 0
\end{array} \right)
\label{a.38}
\end{eqn}

Type {\bf III}$^-$ ($\varepsilon=-1$)

\begin{eqn}
\omega_{ab} &=& \left(
\begin{array}{cccc}
  0     &    -1    & -1    & 0   \\
  1     &     0    &  0    & 0   \\
  1     &     0    &  0    & 0   \\
  0     &     0    &  0    & 0
\end{array} \right)
\label{a.39}
\end{eqn}
The two Casimir invariants vanish for type {\bf III} and yet the
matrix $\omega_{ab}$ is not zero.

 Type {\bf IV} does not exist. Indeed for the case of nilpotency of order 4,
one has $\omega_{ab} l^b=0$,
$\omega_{ab} m^b=l_a$, $\omega_{ab} u^b=m_a$ and $\omega_{ab}
t^b =u_a$. By taking the scalar product of the equation with
$l^a$, one finds $l\cdot l= l\cdot m= l\cdot u=0$. So $l\cdot
t\neq 0$, say $l\cdot t=k$. But then $u\cdot m=m\cdot \omega t=
-l\cdot t \neq 0$ (from the last relations), while the equations
 $\omega_{ab} u^b=m_a$ and the antisymmetry of  $\omega_{ab}
u^b$ imply $u\cdot m=0$. This contradiction shows that type {\bf
IV} is inconsistent.

\newpage

\noindent {\large {\bf A.11 ~~Summary of Results}}

We summarize our results by giving for each type the canonical
form of the Killing vector $(1/2)\omega^{ab} J_{ab}$ and the
corresponding Casimir invariants in a table.

\begin{center}
\begin{tabular}{|r|l|c|c|} \hline
Type     &  Killing vector & $\frac{1}{4} I_1$ & $\frac{1}{4} I_2$ \\ \hline
{\bf I}$_a$    & $b(J_{01}+J_{23}) - a(J_{03}+J_{12}) $ & $b^2-a^2$ &$b^2+a^2$
 \\
{\bf I}$_b$    & $\lambda_1 J_{12} + \lambda_2 J_{03} $ &
		$-\frac{1}{2}(\lambda_1^2 +\lambda_2^2)$ & $\lambda_1
\lambda_2$ \\
{\bf I}$_c$    & $b_1 J_{01} + b_2 J_{23}$ & $\frac{1}{2}(b_1^2 +b_2^2)$
		& $ b_1 b_2$ \\
{\bf II}$_a$   & $\lambda(J_{03} +J_{12}) +J_{01} -J_{02} -J_{13}+J_{23}$ &
 $-\lambda^2$ &           $\lambda^2$ \\
               & or & & \\
&$\lambda (-J_{03} + J_{12}) -J_{13} + J_{23} \ (\lambda \not= 0)$ &
 $-\lambda^2$ & $-\lambda^2$ \\
{\bf II}$_b$   & $(b-1)J_{01} + (b-1)J_{23} +J_{02} -J_{13}$ & $b^2$ &
$b^2$ \\
{\bf III}$^+$  & $-J_{13} + J_{23}$ & 0 & 0 \\
{\bf III}$^-$  & $-J_{01} + J_{02}$ & 0 & 0 \\ \hline
\end{tabular}
\end{center}

\centerline{{\bf Table 1.} Classification of one-parameter subgroups of
 $SO(2,2)$}
\vspace{7mm}

Note that for the second canonical form of type {\bf II}$_a$, valid when
 $\lambda \not= 0$, we have replaced $J_{03}$ by $-J_{03}$ to comply with the
 form given in the text. This amounts to replace $\lambda_2$ by $-\lambda_2$,
 and can be acheived by redefinining $\xi^0$ as $-\xi^0$. This is why the
second Casimir invariant, which is not parity-invariant, changes its sign.

The cases of interest for the black hole are {\bf
I}$_b$, {\bf II}$_a$ and {\bf III}$^+$, for which the
eingenvalues of $\omega_{ab}$, namely $\pm r_{+}/l$ and $\pm r_{-}/l$  are all
 \underline{real}. (These cases exist only because
the signature of the metric is $(--++)$). Type {\bf I}$_b$ (with
$\lambda_1\neq \lambda_2$) describe a general black hole with
$|J|<Ml$, type {\bf II}$_a$ describes an extreme black hole with
non-zero mass, while type {\bf III}$^+$ describes the ground
state with $M=0$. The type becomes more and more special [from four
distinct real roots to one single real root (zero)] as one goes
from the general black hole to the ground state.

It is interesting to notice that if one expresses $r_+$ and
$r_-$ as functions of $J$ and $M$ and goes beyond the extreme
limit $|J|=Ml$, the roots $r_+$ and $r_-$ become complex
conjugates. This strongly suggests that type {\bf I}$_a$
describes the spacetime whose metric is obtained by setting
$|J|>Ml$ in the black hole line element. On the other hand, if
one keeps $|J|<Ml$ and takes $M<0$, the roots $r_+$ and $r_-$
become two different purely imaginary numbers. This strongly
suggests that there is a close relationship between type {\bf
I}$_c$ and the negative mass solutions of \cite{5}.

Finally, on an even more parenthetical note, we mention that for
the Euclidean black hole the group $SO(2,2)$ is replaced by
$SO(3,1)$. In that case the eigenvalues of $\omega^a\,_b$ are of
the form $(a, -a, ib, -ib)$ with real $a$ and $b$. This form may
be obtained from that of type {\bf I}$_b$ above by setting
$M_{Euc}=M$, $J_{Euc}=-iJ$ in the formula (\ref{21}), expressing
the eigenvalues in terms of $M$ and $J$. This is just the
prescription for the (real) Euclidean continuation of the
Minkowskian signature black hole [see, for example\cite{1}.

\vspace{1cm}

\noindent {\Large {\bf Appendix B.~~Smoothness of the Black Hole

                       ~~~~~~~~~~~~~~~~Geometry}}

\renewcommand{\theequation}{B.\arabic{equation}}
\setcounter{equation} 0

This Appendix addresses the question of whether the smoothness
of anti-de Sitter space subsists after the identifications
leading to the black hole are made. That is, we ask whether the
quotient spaces we deal with are Hausdorff manifolds. The
conclusion is that this is so when $J \neq 0$, but when $J=0$
the Hausdorff manifold structure is destroyed at $r=0$.

As discussed by Hawking and Ellis \cite{9},the quotient spaces are Hausdorff
manifolds if and only if the action of the identification
subgroup $H = \{ exp 2\pi k\xi, k\epsilon {\bf Z} \}$ is
properly discontinuous, namely, if the following properties hold,

\begin{description}
 \item [(i)] Each point $Q \epsilon \; adS $ has a neighbourhood
	$U$ such that $(exp 2\pi k\xi)(U) \cap U = \phi$ for all $k
	\epsilon {\bf Z}$, $k \neq 0$; and

 \item [(ii)] If $P,\; Q \; \epsilon \; adS $ do not belong to the
	same orbit of $H$ (i.e., there is no $k\; \epsilon {\bf
	Z}$ such that $(exp 2\pi k \xi)(P) =Q$), then there are
	neighborhoods $B$ and $B'$ of $P$ and $Q$ respectively such
	that $(exp 2\pi k \xi)(B) \cap B' =\phi$ for all $k\;
	\epsilon {\bf Z}$.
\end{description}

To proceed with the analysis we introduce the Euclidean norm
\bb
[(u'-u)^2 + (v'-v)^2 +(x'-x)^2 +(y'-y)^2]^{1/2}
\ee
on {\bf R}$^4$. The norm of the Killing vector
\bb
\xi= \frac{r_+}{l}\left( u\frac{\p}{\p x}+x\frac{\p}{\p u}
\right) - \frac{r_-}{l}\left( v\frac{\p}{\p y} + y\frac{\p}{\p v}
\right)
\ee
is bounded from below by $r_- >0$,
\br
\parallel \xi\cdot\xi\parallel_{E} &=& \left[
\frac{r_+^2}{l^2}(u^2 +x^2) +\frac{r_-^2}{l^2}(v^2 +y^2) \right]
\nonumber \\
&=& \left[ \frac{r_+^2 -r_-^2}{l^2}(u^2 +x^2) +\frac{r_-^2}{l^2}
(u^2 +x^2 +v^2 +y^2) \right]^{1/2} \nonumber \\
&\geq & r_- \;>0 \;\;\; (\mbox{on}\;\; u^2 +v^2 =x^2 +y^2 +l^2)
\label{b.1}
\er

Let $Q_0$ be a point of anti-de Sitter space with coordinates $(u_0, v_0, x_0,
 y_0)$ satisfying $u^2_0 + v^2_0 - x^2_0 - y^2_0 = l^2$.
Its successive images $Q_{n}$ are given by

\br
u_{n} &=& (\cosh \alpha)u_0 + (\sinh \alpha)x_0
\aum \label{b.2.a} \\
x_{n} &=& (\sinh \alpha)u_0 + (\cosh \alpha)x_0
\aum \label{b.2.b} \\
v_{n} &=& (\cosh \beta)v_0 - (\sinh \beta)y_0
\aum \label{b.2.c} \\
y_{n} &=& - (\sinh \beta)v_0 + (\cosh \beta)v_0
\aum \label{b.2.d} \\
\er
with $n\; \epsilon \;{\bf Z}, \;\; \alpha=2\pi r_+/l, \;\; \beta=
2\pi r_-/l$. The Euclidean distance $d_{E}(Q_0,Q_{n})$,
$(n \neq 0)$ between $Q_0$ and $Q_{n}$ is bounded from below by
\bb d_{E}(Q_0,Q_{n}) \geq l\sqrt{2(\cosh \beta -1)} >0, \;(n\neq
0).
\label{b.3}
\ee

Indeed, one has
\br
\lefteqn{(u_{n} -u_0)^2 +(x_{n} -x_0)^2 +(v_{n} -v_0)^2 +(y_{n}
-y_0)^2} \\
&\geq & |(u_{n} -u_0)^2 -(x_{n} -x_0)^2| +|(v_{n} -v_0)^2 -(y_{n}
-y_0)^2| \nonumber \\
&=& 2(\cosh n \alpha -1)|u_0^2 -x_0^2| +
2(\cosh n \beta -1)|v_0^2 -y_0^2| \nonumber \\
&\geq & 2(\cosh \beta -1)[|u_0^2 -x_0^2| +|v_0^2 -y_0^2|] \nonumber \\
&\geq & 2(\cosh \beta -1)|u_0^2 -x_0^2 +v_0^2 -y_0^2| \nonumber \\
&= & 2(\cosh \beta -1)l^2
\er
The bound (B.9) is uniform, i.e., it does not depend on $Q_0$.

Let $P_0$ be another point of anti-de Sitter space with coordinates
$(\bar{u}_0,
 \;\bar{v}_0,
\;\bar{x}_0, \;\bar{y}_0)$. It is easy to see, by using formulas
analogous to (B.3) for $P_0$, that the distance
$d_{E}(P_{n},Q_0)$ between $Q_0$ and the images of $P_0$ goes to
infinity as $n \rightarrow \pm \infty$. Hence, there is a
minimum ``distance of approach" of the orbit of $P_0$ to $Q_0$
(which may be zero if $Q_0 =P_{k}$ for some $k$). That minimum
distance of approach varies continuously if one varies $P_0$ continuously.

Let $U$ be the open ball centered at $Q_0$ with radius
$r<\frac{l}{2} \sqrt{2(\cosh \beta -1)}$. The image of any point
of this ball by $exp 2\pi k\xi$, ($k \neq 0$) cannot be in $U$.
Otherwise the bound (B.3) would be violated. This proves (i).

Now, turn to (ii). Let $P_0$ be a point that is not mapped on
$Q_0$ by any power of $exp 2\pi \xi$. In the open ball $U$,
there can be at most one image of $P_0$. If there were none, by
continuity, the points sufficiently close to $P_0$ will have no
image in $U$ and thus (ii) would hold. So let us assume that
there is one image of $P_0$ in $U$, say $P_{n}$. Let $\tilde{B}$
be an open ball centered at $P_{n}$ and entirely contained in
$U$. All the images of the points in $\tilde{B}$ lie outside
$U$, i.e., $(exp 2\pi k \xi)(\tilde{B}) \cap U =\phi$. Let $B"$
be an open ball centered at $Q_0$ such that $B \cap \tilde{B}
=\phi$. Then $B=(exp -2\pi n\xi)(\tilde{B})$ and $B"$ fulfill
condition (ii).

[For simplicity we have used in this analysis the simpler form
of the Killing vector only appropriate for $|J|<Ml$. One can
easily check that for $|J|=Ml$ there are no fixed points and
that all the orbits go to infinity, just as for $|J|<Ml$. It
then easily follows that the results for $|J|<Ml$ remain valid
for $|J|=Ml$. The details are left to the reader]

The above argument breaks down when there is no angular momentum
because the Killing vector $\xi =\frac{r_+}{l}(u\frac{\p}{\p x}
+x\frac{\p}{\p u})$ vanishes in that case along the line $u=x=0$, which
is thus a line of fixed points. This makes the bound (B.3)
empty. Furthermore, each fixed point is  an accumulation point for
the orbits of the points obeying $u\pm x =0$ and having the same
values of $v$ and $y$. Hence, both (i) and (ii) are violated if
one takes for $Q$ one of the fixed points. The action of the
group is not properly discontinuous. This leads to a singularity
in the manifold structure of the Taub-NUT type.

[This kind of singularity has been discussed in \cite{14}.
Another example of it has been found in \cite{15}. For an
analysis see \cite{9}, where a discussion of identifications
under boosts in two-dimensional Minkowski space is given. To
make contact with that analysis observe that near $r=0$ one can
neglect the cosmological constant. The $SO(2,2)$ group goes then
over to the Poincar\'e group in three dimensions. The
identification Killing vector (\ref{3.15}) becomes then a boost
plus a translation in a transverse direction. It is the presence,
in our case,  of this additional transverse direction which is
responsible for the smooth behavior when $J\neq 0$: the
combination of a boost and a transverse translation does not
have fixed points.]


\begin{center}
\rule{3cm}{.4mm}
\end{center}


\newpage

\centerline{\bf FIGURE CAPTIONS}

\bigskip
\noindent
{\bf Figure 1. \ Poincar\'e Patches}
\begin{enumerate}
\item[(a)] Section with surface $y=0$. The solid lines have $u+x=0,\;\; y=0$.
These
 curves are lightlike and asymptotic to $\lambda = (k+1/2)\pi$. The pattern is
 periodic in $\lambda$.

\item[(b)] Section with surface $x=0$. The solid lines (including the axis
$\lambda
 =0$) have  $u+x =0,\; x=0$ in anti-de Sitter space. The pattern
is again periodic in $\lambda$. As one lets the angle  $\theta$ approach
$\pm \pi/2$, the lines $u+x=0$ become more and more horizontal until they
reach the configuration shown.
\end{enumerate}
\bigskip

\noindent
{\bf Figure 2. \ Regions determined by the norm of} $\xi'$.
\begin{enumerate}
\item[(a)]  Section with surface $y=0$ when $r_- \not= 0$. The solid
lines are the curves $\xi'\cdot \xi' =0$, $y=0$. 
They are timelike. The dotted lines
are the lines $\xi'\cdot\xi' =r_{-}^2$ ($u^2-x^2=0$), 
bounding regions {\bf II} and {\bf III}. The lines formed by dots and segments
 have
$\xi'\cdot \xi'=r_{+}^2$, $y=0$.

\item[(b)] Section with surface $x=0$ when $r_- \not= 0$.
The surface $x=0$ has $\xi'\cdot\xi' > 0$ everywhere when $r_- \not= 0$.  The
 horizontal solid lines are the lines $\xi'\cdot \xi'=r_{-}^2$, $x = 0$. The
 lightlike lines formed by dots and segments have $\xi'\cdot \xi'=r_{+}^2$.
The region
$\xi' \cdot \xi' > 0, x = 0$ splits into disconnected components separated by
 the horizontal lines and containing two regions {\bf I} and two regions {\bf
 II}.

\item[(c)] Section with surface $y=0$ where $r_-=0$. The solid lines
 have $\xi' \cdot \xi' = 0, \ y = 0$. The lines formed by dots and segments
have $\xi' \cdot \xi' = r_+^2$.  The region $\xi'\cdot\xi' > 0$ splits into
 disconnected
components separated by the horizontal lines with each component consisting of
 two regions {\bf II} (and two regions {\bf I}, not seen in this figure since
 they have no intersection with $y=0$). Regions {\bf III} have disappeared.
Note that the Killing vector $\xi'$ is now tangent to the lightlike curves
$u^2-x^2 = 0, y = 0$.
\end{enumerate}
\bigskip
\noindent
{\bf Figure 3. \ Spacetime diagrams for} $J=0$
\begin{enumerate}
\item[(a)] Kruskal diagram,  (b) Penrose diagram.
\end{enumerate}
\bigskip
\noindent
\begin{enumerate}
\item[{\bf Figure 4}]. Penrose diagrams for $J \not= 0$.
(a) Patch $K_-$, \ (b)
 Patch $K_+$, \ (c) Complete diagram obtained by joining an infinite
sequence of
 patches $K_-, K_+$ on the overlap $K$
\noindent
\item[{\bf Figure 5.}] \ Penrose Diagrams for the extreme cases (a)
$M=0=J$, \
 (b) $M= \vert J/l \vert \not= 0.$
\item[{\bf Figure 6.}] \ Penrose diagrams for the maximally extended
 non-extremal spinning black hole $(Ml > \vert J \vert > 0)$, showing also the
 regions beyond the singularity where the Killing vector $\xi$ is timelike.
 Regions {\bf III'} are defined by $- \infty < \xi \cdot \xi < r^2_-$ and
 contain regions {\bf III} \  ($0 < \xi \cdot \xi < r^2_-)$. The metric in
these
 regions is isomorphic to the metric in regions {\bf I} but with the roles of
 $t$ and $\phi$ exchanged. The singularity $r=0$ in {\bf III} corresponds then
 to the stationary surface in {\bf I}. There are closed timelike curves
through
each point in regions {\bf III'}. These closed timelike curves
cross $\xi\cdot\xi=0$.
\item[{\bf Figure 7.}]\ Penrose diagram for a collapsing body in the case
$J=0$.
\end{enumerate}
\end{document}